\documentclass[11pt, oneside]{article}   	% use "amsart" instead of "article" for AMSLaTeX format
\usepackage{geometry}
\usepackage{amsmath}
\numberwithin{equation}{section}

\usepackage{graphicx}
\usepackage{subfigure}
\usepackage{amssymb}
\usepackage{mathrsfs}  
\usepackage[shortlabels]{enumitem}
\usepackage{mathtools}
\usepackage{caption}
\usepackage{cite}
\usepackage{hyperref}
	\hypersetup{
	    colorlinks=true,
	    anchorcolor=black,
	    urlcolor=black,
	    citecolor=red,
	    linkcolor=red}

\def\be{\begin{equation}}
\def\ee{\end{equation}}
\def\bsub{\begin{subequations}}
\def\esub{\end{subequations}}

\title{Renormalization of Schr\"odinger equation for potentials with inverse-square singularities: Generalized Trigonometric P\"oschl-Teller model}
\author{
U. Camara da Silva\thanks{ulysses.silva@ufes.br}
\\
{\small 	Universidade Federal do Esp\'irito Santo} -- 		
{\small		Departamento de F\'isica} \\
{\small	Av. Fernando Ferrari, Goiabeiras, 
		29075-900, 
		Vit\'oria-ES, Brasil}
}
\date{}							% Activate to display a given date or no date

\begin{document}
\pagenumbering{gobble}

\maketitle
\begin{abstract}
We introduce a renormalization procedure necessary for the complete description of the energy spectra of a one-dimensional stationary Schr\"odinger equation with a potential that exhibits inverse-square singularities. We apply and extend the methods introduced in our recent paper on the hyperbolic P\"oschl-Teller potential (with a single singularity) to its trigonometric version. This potential, defined between two singularities, is analyzed across the entire bidimensional coupling space. The fact that the trigonometric P\"oschl-Teller potential is supersymmetric and shape-invariant simplifies the analysis and eliminates the need for self-adjoint extensions in certain coupling regions. However, if at least one coupling is strongly attractive, the renormalization is essential to construct a discrete energy spectrum family of one or two parameters. We also investigate the features of a singular symmetric double well obtained by extending the range of the trigonometric P\"oschl-Teller potential. It has a non-degenerate energy spectrum and eigenstates with well-defined parity.

\end{abstract}

\tableofcontents

\section{Introduction}

Physical systems whose properties are described by stationary Schr\"odinger equations with potentials exhibiting inverse-square singularities present a challenging task. Depending on the values of the parameters, the proper definition of the Schr\"odinger operator as a self-adjoint one and the proof of the existence of normalizable solutions require the application and further development of analytical methods, including appropriate regularization, the introduction of dynamical scales, and a renormalization procedure. This approach provides a systematic description of the changes in the energy spectrum properties and the corresponding eigenstates across different regions of the coupling space.

The present paper provides a complete solution to all these problems for the generalized trigonometric P\"oschl-Teller (TPT) potential
\begin{eqnarray}
\frac{2m}{\hbar^2}V(x)\equiv\mathcal{V}(x)=\alpha^2\left(\frac{g_s}{\sin^2(\alpha x)}+\frac{g_c}{\cos^2(\alpha x)}\right), \ 0<x<\frac{\pi}{2\alpha},\label{V_PT_trig}
\end{eqnarray}
where $\alpha$ has dimensions of inverse length, and the couplings $g_s$ and $g_c$ are dimensionless. It represents a particle confined to a finite spatial interval defined by a potential with inverse-square singularities at both ends \cite{Poschl:1933zz}. 

The generalized TPT potential describes the linear fluctuations around $AdS_d$ solutions of the Einstein gravity. The above potential has been investigated in the well-known paper by Wald and Ishibashi  \cite{Ishibashi:2004wx}. They have shown that the mentioned linear fluctuations in $AdS_d$ space in global coordinates are all described by the TPT  potential (\ref{V_PT_trig}), with specific parametrization of the couplings $g_s$ and $g_c$  in terms of the space-time dimension $d$, the mass, and the spin of the fields. The generalized hyperbolic version of our TPT case, \cite{Poschl:1933zz}, usually represents linear fluctuations around the so-called $d=2+1$ BTZ black hole \cite{PhysRevD63124015}. The appearance of all these potentials and even more general ones involving inverse-squared singularities, i.e., of the Calogero or conformal type \cite{Calogero:1970nt,deAlfaro:1976vlx}, is a rather common phenomenon in gauge/gravity correspondence problems when the bulk gravity solutions are representing one or, say, two (asymptotically) $AdS$ space-time regions. Equation (\ref{V_PT_trig}) exhibits exactly this behavior near the two singularities - $V(x)\approx g_s/x^2$ as $x\rightarrow0$ and $V(x)\approx g_c/(\frac{\pi}{2\alpha}-x)^2$ as $x\rightarrow \frac{\pi}{2\alpha}$ - characterizing asymptotic conformal symmetries at both extremes.

The main result of the present paper concerns the construction of families of energy eigenstates for the generalized TPT  potential across the entire $g_s$-$g_c$ coupling space, including regions where the Hamiltonian does not have a self-adjoint extension, \cite{Ishibashi:2004wx}. Instead of following a self-adjoint extension approach, \cite{Gitman_Book,Gitman,Falomir:2005yw,ConfQM}, we will address the problem through a renormalization process, analogous to what has been proposed for the conformal potential \cite{renor_orig,Renor_Inver_Squa,Camblong:2003mb,RG_limit_c,renor_Group_Limit_Cycle,Bawin:2003dm,U_AB} and with certain modifications to the generalized hyperbolic P\"oschl-Teller (HPT) potential  considered in our recent paper \cite{CAMARADASILVA2024169549z}. The renormalized model possesses a discrete energy spectrum, bounded from below or not, that is well-defined for any pair $(g_s,g_c)$ of couplings. The key difference between the TPT and the HPT is that here there are two independent renormalization processes, one for each singularity, resulting in families of one- or two-parameter energy eigenfunctions, where the two possible scales arising via ``dimensional transmutation'', $L_{\nu_s}$ and $L_{\nu_c}$, spontaneously break the asymptotic conformal symmetries. When both scales are at fixed points, i.e., for $L_{\nu_s}=0$ or $\infty$ and $L_{\nu_c}=0$ or $\infty$, there is no spontaneous breaking of the asymptotic conformal symmetries at all. The TPT potential is supersymmetric (SUSY) and shape-invariant for $g_s,g_c\ge-1/4$,\footnote{In the context of asymptotically $AdS_d$ spaces and holography \cite{Maldacena:1999uc,Witten_1998}, this is  the famous BF-bound, \cite{Breitenlohner:1982uc,Breitenlohner:1982ua}.} and as we will argue the sufficient condition is to take $L_{\nu_s}=L_{\nu_c}=0$. Thus, spontaneous breaking of the asymptotic conformal symmetry occurs only when the SUSY is no longer valid.

Finally, the last topic to be addressed in this article is extending the potential to the domain $-\frac{\pi}{2\alpha}<x<\frac{\pi}{2\alpha}$, turning it into a symmetric well with a singularity at the center. When the coupling $g_s$ lies in the interval $-1/4\le g_s<3/4$, renormalization allows us to construct one-parameter families of non-degenerate eigenstates with well-defined parity. This result is analogous to that found in \cite{CAMARADASILVA2024169549z} for the hyperbolic P\"oschl-Teller (HPT) potential, demonstrating the versatility of the renormalization approach for potentials with inverse-square singularities.

The structure of the paper is as follows. In Section \ref{sec:2}, we present the general solution of the model and its features near the singularities, highlighting the different behaviors within the $g_s-g_c$ coupling space and the possible ambiguities in defining the boundary conditions. Section \ref{sec:3} is dedicated to the regions where the TPT potential is SUSY shape-invariant, demonstrating how this property allows us to uniquely determine the boundary conditions, even in the regions usually considered problematic. In Section \ref{sec:4}, the renormalization procedure is applied to make the boundary conditions well-defined throughout the entire $g_s-g_c$ plane by introducing scales that break the asymptotic conformal symmetry of the TPT potential. In Section \ref{sec:espectro}, the renormalized spectrum is systematically determined in all subregions of the coupling space, including deep-bound states that exhibit the \emph{Efimov Rule}, \cite{RG_limit_c,Efimov}. In Section \ref{sec:duplo}, the potential is extended to the larger domain $-\frac{\pi}{2\alpha}<x<\frac{\pi}{2\alpha}$, thus creating a symmetric well, and we show that renormalization combined with the imposition of well-defined parity eigenstates can generate a well-defined non-degenerate spectrum. Finally, in Section \ref{sec:conc}, we present our final discussions. The renormalization along the critical line, $g_s=-1/4$, is provided in Appendix \ref{sec:app}.

%%%%%%%%%%%%%%%%%%%%%%%%%%%%%%%%%%%%%%%%%%%%%%%%%%%%%%%%%
\section{General solution and boundary conditions}\label{sec:2}

\begin{figure}[ht] 
\centering
\subfigure[]{
\includegraphics[scale=0.35]{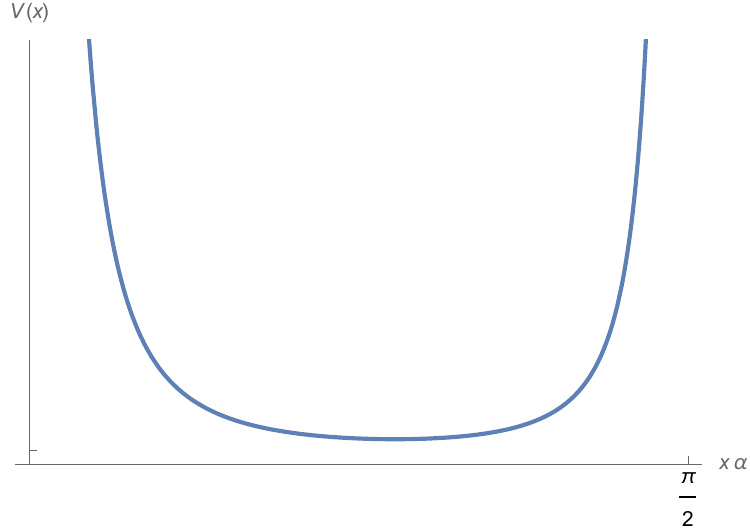}
\label{fig:V_1}
}
\subfigure[]{
\includegraphics[scale=0.35]{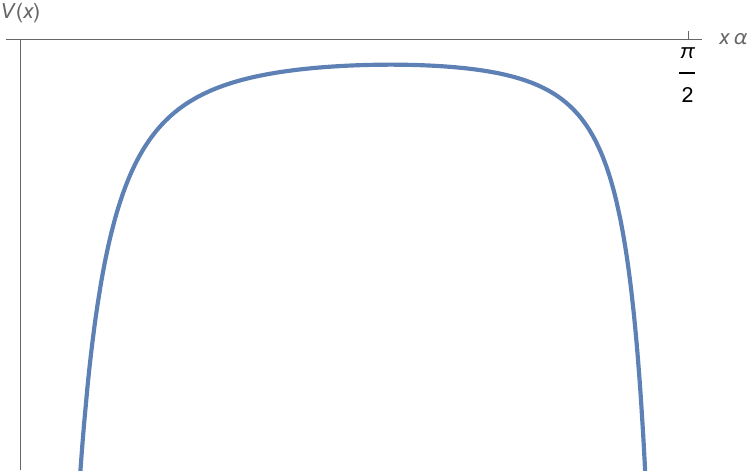}
\label{fig:V_2}
}
\subfigure[]{
\includegraphics[scale=0.35]{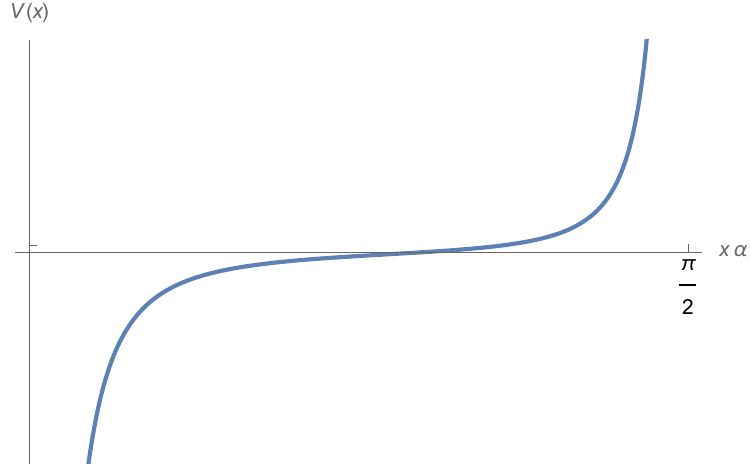}
\label{fig:V_3}
}
\caption{Possible forms of the potential (\ref{V_PT_trig}): (a) $g_s,g_c>0$ (b) $g_s,g_c<0$; (c) $g_s<0$ and $g_c>0$.
}
\label{fig:V_PT}
\end{figure}

The TPT potential, given by equation (\ref{V_PT_trig}), has a minimum (maximum) when $g_s,g_c>0$ ($g_s,g_c<0$). In the case where the couplings have opposite signs, there are no extrema. The three qualitatively different curves are represented in Figure \ref{fig:V_PT}.

Our goal is to study the square-integrable solutions of the stationary Schr\"odinger equation for the potential (\ref{V_PT_trig}), that is\footnote{The definitions used are equivalent to choosing natural units $\hbar=2m=1$ with $V(x)\rightarrow\mathcal{V}(x)$ and $E\rightarrow k^2$.}:
\begin{eqnarray}
\frac{d^2\psi(x)}{dx^2}+\left(k^2-\mathcal V(x)\right)\psi(x)=0, \ k=\sqrt{\frac{2mE}{\hbar^2}}.\label{Sch_x}
\end{eqnarray}
Since the system exists in a finite interval, the energy spectrum, $E$, will always be discrete. However, the eigenstates are sensitive to the sign of the energy: when it is positive, we have $k>0$, while for negative energies, $k=i\kappa$, with $\kappa>0$. The general solution can be written in terms of hypergeometric functions \cite{NIST:DLMF}.
\begin{eqnarray}
&&\psi_{k;\nu_s,\nu_c}(x)=\frac{A_{\nu_s}(k)}{\alpha^{1/2+\nu_s}}\left(\tan(\alpha x)\right)^{\frac{1}{2}+\nu_s}\left(\cos(\alpha x)\right)^{-\frac{k}{\alpha}}F\left(\nu_s,\nu_c,x\right)+\nonumber\\
&&\frac{B_{\nu_s}(k)}{\alpha^{1/2-\nu_s}}\left(\tan(\alpha x)\right)^{\frac{1}{2}-\nu_s}\left(\cos(\alpha x)\right)^{-\frac{k}{\alpha}}F\left(-\nu_s,\nu_c,x\right),\label{sol_G}\\
&&F\left(\pm\nu_s,\nu_c,x\right)\equiv\,{}_2F_1\left(\frac{1}{2}\pm\frac{\nu_s}{2}-\frac{\nu_c}{2}+\frac{k}{2\alpha},\frac{1}{2}\pm\frac{\nu_s}{2}+\frac{\nu_c}{2}+\frac{k}{2\alpha};1\pm\nu_s;-\tan^2(\alpha x)\right),\nonumber
\end{eqnarray}
%%%%%%%%%%%%%%%%%%%%%%%%%%%%%%%%%%%%%%%%%%%%%%%%%%%%%%%%%%
%%%%%%%%%%%%%%%%%%%%%%%%%%%%%%%%%%%%%%%%%%%%%%%%%%%%%%%%
with
\begin{eqnarray}
\nu_s=\sqrt{\frac{1}{4}+g_s}, \ \ \nu_c=\sqrt{\frac{1}{4}+g_c},\label{acop}
\end{eqnarray}
and the parameters $A_{\nu_s}(k)$ and $B_{\nu_s}(k)$ are the coefficients of the two linearly independent solutions of the differential equation chosen in that way to simplify the behavior of the eigenfunction in the vicinity of the singularity $x=0$, see below. Now, we will present the asymptotic forms of the general solution near each singularity to analyze their dependence on the values of the couplings ($g_s,g_c$). Starting with the asymptotic behavior for $\alpha x\ll1$, for $g_s>-1/4$ ($\nu_s>0$), we have
\begin{eqnarray}
\psi_{k;\nu_s,\nu_c}(x)\approx x^{1/2}\left(A_{\nu_s}(k)x^{\nu_s}+B_{\nu_s}(k)x^{-\nu_s}\right).\label{x_s}
\end{eqnarray}
If $g_s<-1/4$, the parameter $\nu_s$ becomes purely imaginary, $\nu_s=i|\nu_s|=i\sqrt{|g_s|-1/4}$. For one-dimensional bound states, the probability current density must vanish; otherwise, the Hamiltonian could not be self-adjoint, so the energy eigenstates are real functions. Therefore, without loss of generality, we can define
\begin{eqnarray}
A_{i|\nu_s|}(k)=\frac{|C_{|\nu_s|}(k)|}{2i}k^{i|\nu_s|}e^{i\zeta_{|\nu_s|}(k)}=B_{i|\nu_s|}^*(k)\label{A_B_theta}
\end{eqnarray}
and the asymptotic behavior becomes
\begin{eqnarray}
\psi_{k;\nu_s,\nu_c}(x)\approx |C_{|\nu_s|}(k)|x^{1/2}\sin\left(|\nu_s|\ln(kx)+\zeta_{|\nu_s|}(k)\right).\label{x_<}
\end{eqnarray}

The function (\ref{sol_G}) behavior for $\pi/2-\alpha x\ll1$ has the same structure as the previous asymptotic form, with the dominant term related to the coupling $g_c$. Its form is
\begin{eqnarray}
\psi_{k;\nu_s,\nu_c}(x)\approx A_{\nu_c}(k)\left(\frac{\pi}{2\alpha}-x\right)^{1/2+\nu_c}+B_{\nu_c}(k)\left(\frac{\pi}{2\alpha}-x\right)^{1/2-\nu_c}, \ g_c>-1/4,\label{x_c}
\end{eqnarray}
and
\begin{eqnarray}
&&\psi_{k;\nu_s,\nu_c}(x)\approx|C_{\nu_c}(k)|x^{1/2}\sin\left(|\nu_c|\ln(kx)+\zeta_{|\nu_c|}(k)\right),\label{x_c_2}\\
&&A_{i|\nu_c|}(k)=\frac{|C_{|\nu_c|}(k)|}{2i}k^{i|\nu_c|}e^{i\zeta_{|\nu_c|}(k)}=B_{i|\nu_c|}^*(k), \ g_c<-1/4,\nonumber
\end{eqnarray}
where
\begin{eqnarray}
\frac{A_{\nu_c}(k)}{\alpha^{\nu_c}}=&&\frac{A_{\nu_s}(k)}{\alpha^{\nu_s}}\frac{\Gamma(-\nu_c)\Gamma(1+\nu_s)}{\Gamma\left(\frac{1}{2}+\frac{\nu_s}{2}-\frac{\nu_c}{2}+\frac{k}{2\alpha}\right)\Gamma\left(\frac{1}{2}+\frac{\nu_s}{2}-\frac{\nu_c}{2}-\frac{k}{2\alpha}\right)}+\nonumber\\
&&\frac{B_{\nu_s}(k)}{\alpha^{-\nu_s}}\frac{\Gamma(-\nu_c)\Gamma(1-\nu_s)}{\Gamma\left(\frac{1}{2}-\frac{\nu_s}{2}-\frac{\nu_c}{2}+\frac{k}{2\alpha}\right)\Gamma\left(\frac{1}{2}-\frac{\nu_s}{2}-\frac{\nu_c}{2}-\frac{k}{2\alpha}\right)},\nonumber\\
\frac{B_{\nu_c}(k)}{\alpha^{-\nu_c}}=&&\frac{A_{\nu_s}(k)}{\alpha^{\nu_s}}\frac{\Gamma(\nu_c)\Gamma(1+\nu_s)}{\Gamma\left(\frac{1}{2}+\frac{\nu_s}{2}+\frac{\nu_c}{2}+\frac{k}{2\alpha}\right)\Gamma\left(\frac{1}{2}+\frac{\nu_s}{2}+\frac{\nu_c}{2}-\frac{k}{2\alpha}\right)}+\nonumber\\
&&\frac{B_{\nu_s}(k)}{\alpha^{-\nu_s}}\frac{\Gamma(\nu_c)\Gamma(1-\nu_s)}{\Gamma\left(\frac{1}{2}-\frac{\nu_s}{2}+\frac{\nu_c}{2}+\frac{k}{2\alpha}\right)\Gamma\left(\frac{1}{2}-\frac{\nu_s}{2}+\frac{\nu_c}{2}-\frac{k}{2\alpha}\right)}.\label{master}
\end{eqnarray}
Or alternatively
\begin{eqnarray}
\frac{\alpha^{2_c} B_{\nu_c}(k)}{A_{\nu_c}(k)}=\frac{
\frac{\Gamma(\nu_c)\Gamma(1-\nu_s)}{\Gamma\left(\frac{1}{2}-\frac{\nu_s}{2}+\frac{\nu_c}{2}+\frac{k}{2\alpha}\right)\Gamma\left(\frac{1}{2}-\frac{\nu_s}{2}+\frac{\nu_c}{2}-\frac{k}{2\alpha}\right)}\frac{\alpha^{2\nu_s}B_{\nu_s}(k)}{A_{\nu_s}(k)}+\frac{\Gamma(\nu_c)\Gamma(1+\nu_s)}{\Gamma\left(\frac{1}{2}+\frac{\nu_s}{2}+\frac{\nu_c}{2}+\frac{k}{2\alpha}\right)\Gamma\left(\frac{1}{2}+\frac{\nu_s}{2}+\frac{\nu_c}{2}-\frac{k}{2\alpha}\right)}}{
\frac{\Gamma(-\nu_c)\Gamma(1-\nu_s)}{\Gamma\left(\frac{1}{2}-\frac{\nu_s}{2}-\frac{\nu_c}{2}+\frac{k}{2\alpha}\right)\Gamma\left(\frac{1}{2}-\frac{\nu_s}{2}-\frac{\nu_c}{2}-\frac{k}{2\alpha}\right)}\frac{\alpha^{2\nu_s}B_{\nu_s}(k)}{A_{\nu_s}(k)}+\frac{\Gamma(-\nu_c)\Gamma(1+\nu_s)}{\Gamma\left(\frac{1}{2}+\frac{\nu_s}{2}-\frac{\nu_c}{2}+\frac{k}{2\alpha}\right)\Gamma\left(\frac{1}{2}+\frac{\nu_s}{2}-\frac{\nu_c}{2}-\frac{k}{2\alpha}\right)}}.\label{master_2}
\end{eqnarray}
An interesting observation is that the coefficients are related by a \emph{Möbius-type} transformation \cite{mobius}:
\begin{eqnarray}
\frac{\alpha^{2\nu_c} B_{\nu_c}(k)}{A_{\nu_c}(k)}=\frac{\mathcal{M}_1(k)\frac{\alpha^{2_s} B_{\nu_s}(k)}{A_{\nu_s}(k)}-\mathcal{M}_2(k)}{-\mathcal{M}_3(k)\frac{\alpha^{2_c} B_{\nu_s}(k)}{A_{\nu_s}(k)}+\mathcal{M}_4(k)}, \ \mathcal{M}_1(k)\mathcal{M}_4(k)-\mathcal{M}_2(k)\mathcal{M}_3(k)=-\frac{\nu_s}{\nu_c}.\label{M_}
\end{eqnarray}
The non-zero ``determinant'' (the case $\nu_s=0$ is an exception treated in the Appendix \ref{sec:app}) ensures that we can invert the above relationship, and the ratio $\alpha^{2\nu_c}B_{\nu_c}/A_{\nu_c}$ cannot be fixed independently of the ratio $\alpha^{2\nu_c}B_{\nu_s}/A_{\nu_s}$ - if it were zero, we would have fixed the ratio $\alpha^{2\nu_c}B_{\nu_c}/A_{\nu_c}$ as $-\mathcal{M}_2(k)/\mathcal{M}_4(k)$ and no renormalization would have been necessary.

\subsection{Coupling space}

The energy eigenstate of the model must be square-integrable, that is
\begin{eqnarray}
\int_0^{\frac{\pi}{2\alpha}}\left|\psi_{k;\nu_s,\nu_c}(x)\right|^2dx<\infty.\label{quad_int}
\end{eqnarray}
From the asymptotic behaviors (\ref{x_s})-(\ref{x_c}), we have three possibilities for each coupling. The three regions are discussed in detail in \cite{CAMARADASILVA2024169549z} for the generalized hyperbolic PT potential. Here, we will briefly describe them for both $g=g_s$ and $g=g_c$. 
\begin{description}
\item[\emph{i)}] \emph{Strong-repulsive coupling} - $g>3/4$ ($\nu>1$): The solution with asymptotic behavior $\sim |x-x_s|^{1/2-\nu}$ ($x_s=0,\pi/2\alpha$) is not square-integrable. Therefore, the condition (\ref{quad_int}) fixes the constant $B_{\nu}(k)=0$.
 
\item[\emph{ii)}] \emph{Weak-medium coupling} - $-1/4<g<3/4$ ($0<\nu<1$): Both solutions are square-integrable, regardless of whether the interaction is attractive ($g<0$) or repulsive ($g>0$), \cite{renor_m_w_c}. The condition (\ref{quad_int}) does not impose any restrictions, and, a priori, there is no clear criterion to relate $A_{\nu}(k)$ and $B_{\nu}(k)$. This fact is related to the Hamiltonian (\ref{Sch_x}) not being self-adjoint in this case, although it has a bounded below self-adjoint extension, \cite{Ishibashi:2004wx}.

\item[\emph{iii)}] \emph{Strong-attractive coupling} - $g<-1/4$, $\nu\rightarrow i|\nu|$: The eigenstate strongly oscillates near the singularity without converging to a defined value. This case violates the famous BF bound that delimits the stability of fields in AdS$_d$ space \cite{Breitenlohner:1982uc,Breitenlohner:1982ua}, since, in addition to the Hamiltonian not being self-adjoint, it is unbounded below and thus has no positive self-adjoint extension ($k^2>0$), \cite{Ishibashi:2004wx}. On the other hand, the renormalization performed in Section \ref{sec:4} will allow the determination of a well-defined spectrum, including states with negative energy.
\end{description}

\begin{figure}[ht] 
\centering
\subfigure[]{
\includegraphics[scale=0.25]{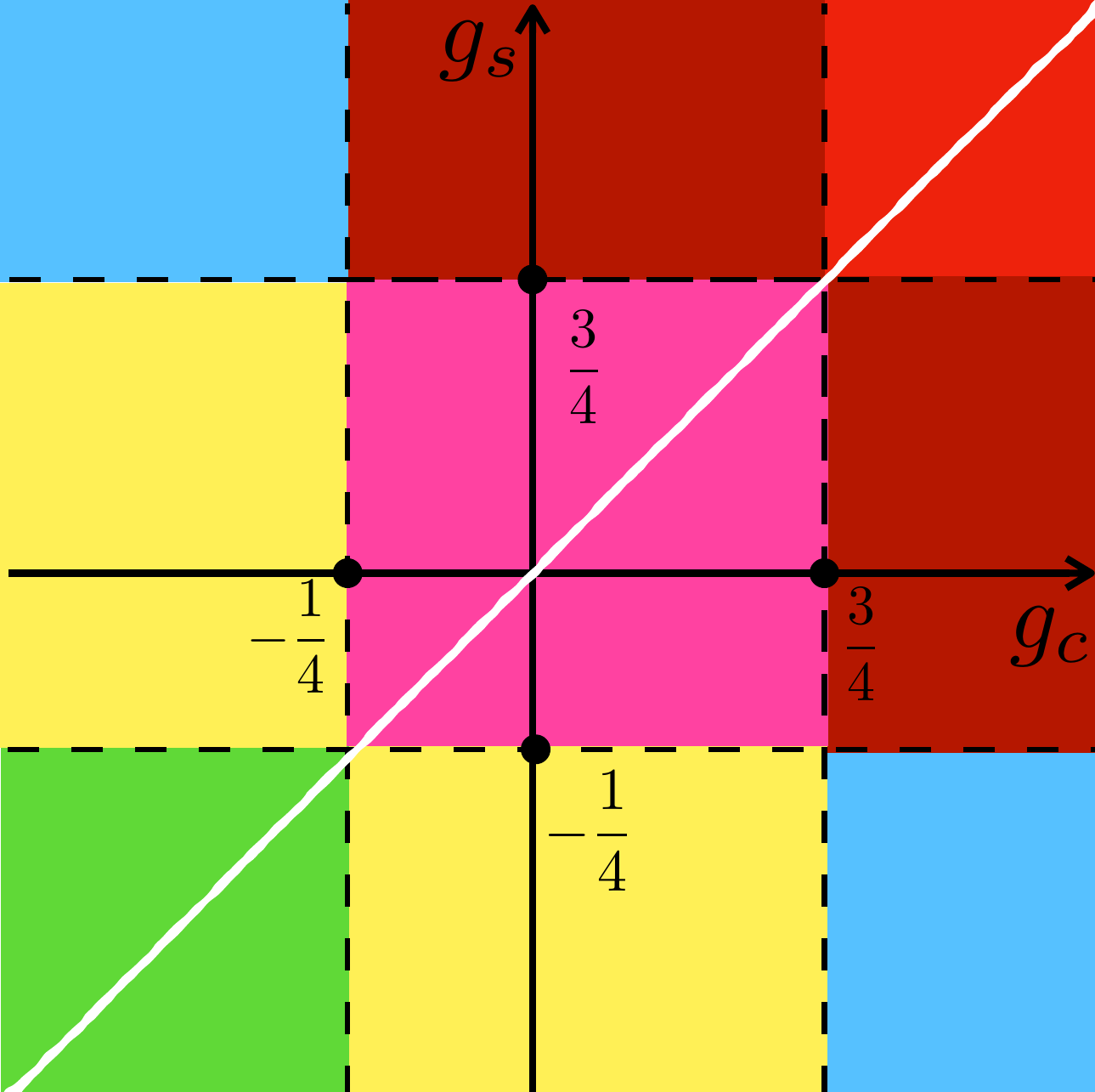}
\label{fig:V_1_2}
}
\caption{Coupling space diagram. The white line $g_s=g_c$ separates the two symmetric half-planes. Red region: strong-repulsive and strong-repulsive. Brown region: strong-repulsive and weak-medium. Blue region: strong-repulsive and strong-attractive. Pink region: weak-medium and weak-medium. Yellow region: weak-medium and strong-attractive. Green region: strong-repulsive and strong-repulsive.}
\label{fig:esp_g}
\end{figure}

Since we have two independent couplings, the two-dimensional coupling space $(g_s,g_c)$ has nine parts. However, both couplings are attached to square-inverse singularities, so the plane is symmetric along the line $g_s=g_c$, with only six physically distinct regions. In Figure \ref{fig:esp_g}, we represent the coupling space, with each physically distinct region described by a different color.
%%%%%%%%%%%%%%%%%%%%%%%%%%%%%%%%%
%%%%%%%%%%%%%%%%%%%%%%%%%%%%%%%%%%%%%%%

Before discussing how boundary conditions relate these coefficients, we note that the TPT potential is SUSY and shape-invariant when $g_c,g_s>-1/4$. We will show that this property is sufficient to determine the boundary conditions.

\subsection{Supersymmetry and boundary conditions}\label{sec:3}
A Hamiltonian is SUSY when it can be rewritten as a product of conjugate operators, \cite{Infeld:1951mw,Cooper:1994eh}.
\begin{eqnarray}
\frac{2m}{\hbar^2}\hat H\equiv\hat{\mathcal{H}}_1=\hat{\mathcal Q}^{\dagger}\hat{\mathcal Q}+k_0^2, \, k_0^2=\frac{2mE_0}{\hbar^2}.\label{SUSY}
\end{eqnarray}
Here, $E_0$ is the ground state energy, and
\begin{eqnarray}
\hat{\mathcal Q}=\frac{d}{dx}+\mathcal{W}(x), \ \ \hat{\mathcal Q}^{\dagger}=-\frac{d}{dx}+\mathcal{W}(x).
\end{eqnarray}
The \emph{superpotential} $\mathcal{W}(x)$ is a real function related to the potential through the identity
\begin{eqnarray}
\mathcal{V}_1(x)=\mathcal{W}^2(x)-\frac{d\mathcal{W}(x)}{dx}+k_0^2.\label{W_TPT}
\end{eqnarray}
Defining the \emph{superpotential} as
\begin{eqnarray}
\mathcal W(x)=-\alpha\left(\nu_s+\frac{1}{2}\right)\cot(\alpha x)+\alpha\left(\nu_c+\frac{1}{2}\right)\tan(\alpha x),\label{W_SUSY}
\end{eqnarray}
we have
\begin{eqnarray}
&&\mathcal{V}_1(x)=\left(\nu_s^2-\frac{1}{4}\right)\frac{\alpha^2}{\sin^2(\alpha x)}+\left(\nu_c^2-\frac{1}{4}\right)\frac{\alpha^2}{\cos^2(\alpha x)},\label{V_1}\\
&&k_0^2=-\alpha^2(\nu_s+\nu_c+1)^2,\nonumber
\end{eqnarray}
exactly the potential (\ref{V_PT_trig}), see equation (\ref{acop}).
A SUSY Hamiltonian has a partner defined by reversing the order of the operators $\hat{\mathcal{Q}}$ and $\hat{\mathcal{Q}}^{\dagger}$, that is
\begin{eqnarray}
\hat{\mathcal{H}}_2=\hat{\mathcal{Q}}\hat{\mathcal{Q}}^{\dagger}+k_0^2=-\frac{d^2}{dx^2}+\mathcal{V}_2(x)=-\frac{d^2}{dx^2}+\mathcal{W}^2(x)+\frac{d\mathcal{W}(x)}{dx}+k_0^2.
\end{eqnarray}
A crucial feature of the TPT is that it is \emph{shape invariant}. This means that $\hat{\mathcal{H}}_2$ has the same form as $\hat{\mathcal{H}}_1$, but with shifted parameters. Indeed, a direct calculation shows that the partner potential associated with the superpotential (\ref{W_SUSY}) is
\begin{eqnarray}
\mathcal{V}_2(x)=\left((\nu_s+1)^2-\frac{1}{4}\right)\frac{\alpha^2}{\sin^2(\alpha x)}+\left((\nu_c+1)^2-\frac{1}{4}\right)\frac{\alpha^2}{\cos^2(\alpha x)},
\end{eqnarray}
which is the potential (\ref{V_1}) with the translations $\nu_s\rightarrow\nu_s+1$ e $\nu_c\rightarrow\nu_c+1$ and $\nu_c\rightarrow\nu_c+1$. Shape invariance can be used as a criterion to fix the boundary conditions of the problem, \cite{Lima:2019xzg}. If the couplings are initially in the pink or brown regions of Figure \ref{fig:esp_g}, the partner model must necessarily be in the red region, where $B_{\nu_s+1}(k)=B_{\nu_c+1}(k)=0$. On the other hand, SUSY also relates the eigenfunctions through the action of $\hat{\mathcal{Q}}^{\dagger}$ ($\hat{\mathcal{Q}}$), for example
\begin{eqnarray}
\psi_{k;\nu_s+1,\nu_c+1}(x)\propto\hat{\mathcal{Q}}\psi_{k;\nu_s,\nu_c}(x).
\end{eqnarray}
Using this result for the solution near the singularities, $x_0=0,\frac{\pi}{2\alpha}$, where $\mathcal{W}(x)\approx-(\nu+1/2)/(x-x_0)$, the solution with the asymptotic behavior $B_{\nu}(k)|x-x_0|^{\frac{1}{2}-\nu}$ leads to the undesirable (non-square-integrable) behavior $\sim |x-x_0|^{\frac{1}{2}-(\nu+1)}$, that is
\begin{eqnarray*}
B_{\nu+1}(k)\propto B_{\nu}(k).
\end{eqnarray*}
Since $B_{\nu+1}(k)=0$ for the eigenfunction to be square-integrable, we have $B_{\nu}(k)=0$. This argument holds for both singularities. The conclusion is that the \emph{shape invariance} of the potential fixes
\begin{eqnarray}
B_{\nu_s}(k)=B_{\nu_c}(k)=0,\label{B=0}
\end{eqnarray}
in the red, pink, and brown regions of Figure \ref{fig:esp_g}. In these cases, starting from $k_0$ and the \emph{shape invariance} of the potential, the entire spectrum is determined trivially as
\begin{eqnarray}
k_n^2=\frac{2m E_n}{\hbar^2}=\alpha^2\left(\nu_s+\nu_c+2n+1\right)^2, n\in\mathbb{Z}_{\ge0}.\label{E_lig_trig}
\end{eqnarray}
We finish this topic by stating that \emph{shape invariance} is a sufficient, but not necessary, condition for the validity of Eq. (\ref{B=0}). As discussed in reference \cite{Lima:2019xzg}, the only requirement is that the potential be \emph{supersymmetric}, with an inverse-square type singularity and coupling in the \emph{weak-medium} region. It ensures that the partner-potential will also exhibit an inverse-square singularity, but in the \emph{strong-repulsive} coupling region.
%%%%%%%%%%%%%%%%%%%%%%%%%%%%%%%%%%%%%%%%%%%%%%%%%%%%%%%%
\begin{figure}[ht] 
\centering
\subfigure[]{
\includegraphics[scale=0.25]{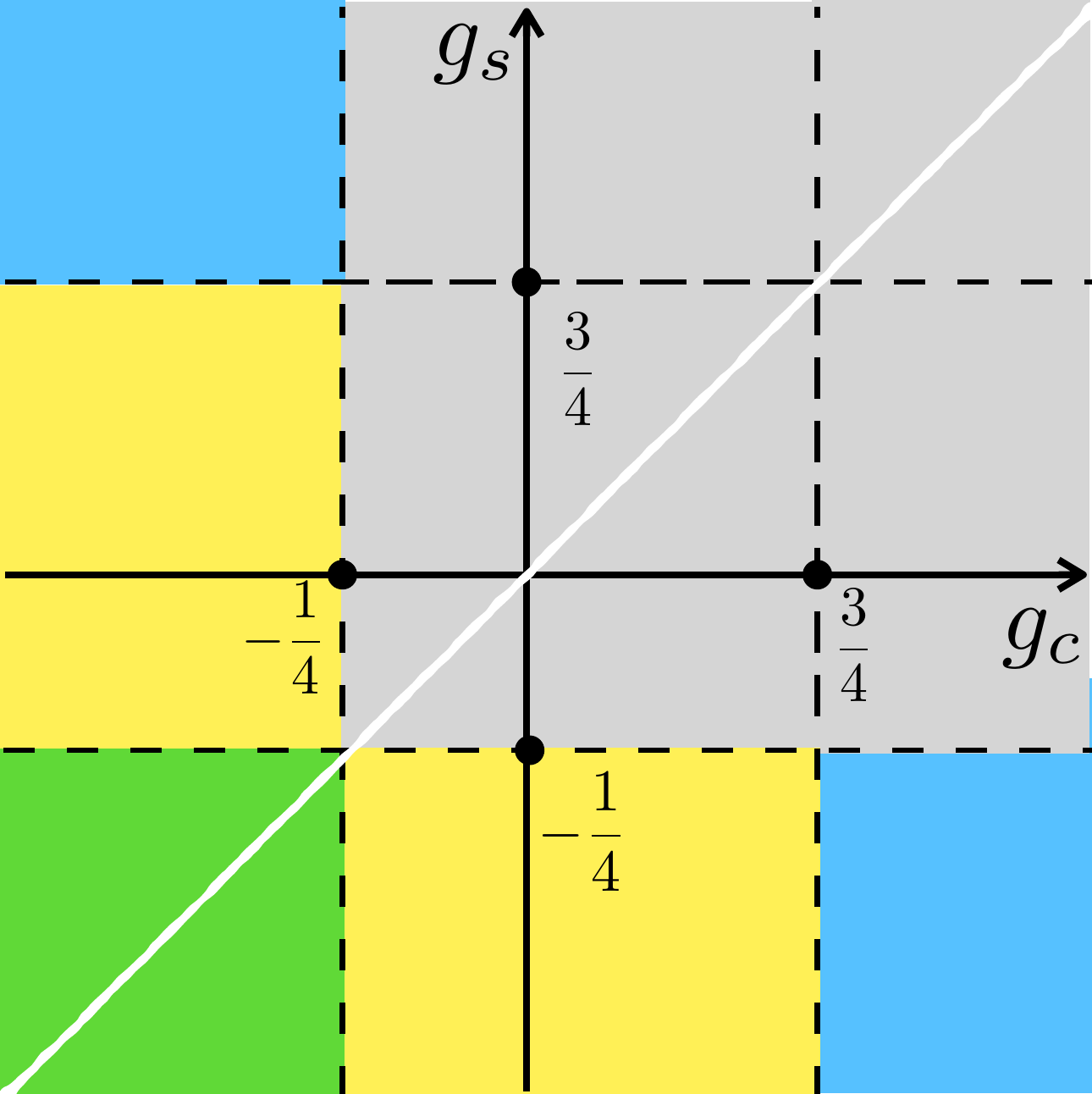}
\label{fig:V_3_2}
}
\caption{Updated coupling space diagram after the use of supersymmetry. Gray region: boundary conditions are well-defined. In the other sectors, renormalization is necessary.
}
\label{fig:esp_g_2}
\end{figure}
%%%%%%%%%%%%%%%%%%%%%%%%%%%%%%%%%%%%%%%%%%%%%%%%%%%%%%%

Where the TPT potential is SUSY, the boundary conditions are fixed; however, SUSY is not valid across the entire coupling space - as we will explain below. In the blue, yellow, and green regions, at least one of the couplings is less than $-1/4$, and consequently, its corresponding $\nu=i|\nu|$. As a result, the superpotential (\ref{W_SUSY}) is no longer a real function; the factorization, formally, still occurs, and the potential $\mathcal{V}(x)$ remains real. A priori, this is not a problem, since there are SUSY models with complex \emph{superpotentials} (and even complex potentials) and real energy spectrum, \cite{ANDRIANOV:1999aa}. On the other hand, in our case, the application of $\hat{\mathcal Q}$ and $\hat{\mathcal Q}^{\dagger}$ (which is not more the adjoint of $\hat{\mathcal{Q}}$) maps different eigenstates with $i|\nu|\rightarrow i(|\nu|+1)$ and complex energy spectrum, both belonging to the \emph{strong-attractive} regime where the strong oscillations of both solutions near the singularity prevent the selection of one. As a consequence, it's not possible to fix a ground state equation, $\hat{\mathcal{Q}}\psi_{0;|\nu|}(x)=0$, so SUSY is broken. Without supersymmetry, the ambiguity remains. In Figure \ref{fig:esp_g_2}, we have the updated map of the coupling space regions. In the gray zone, the boundary conditions are $B_{\nu_s}=B_{\nu_c}=0$, and the spectrum is given by (\ref{E_lig_trig}). In the next section, we will review the method that allows defining appropriate boundary conditions in the other regions of the coupling space through renormalization.

\section{The renormalization of the model}\label{sec:4}

A way to make the boundary conditions well-defined at points where the potential has inverse-square singularities is through renormalization \cite{renor_orig}. This approach is well-known for the exact conformal potential \cite{renor_orig,Renor_Inver_Squa,renor_Group_Limit_Cycle,renor_m_w_c,Gitman_Book}, with results equivalent to those of a self-adjoint extension, and has also been applied to the hyperbolic P\"oschl-Teller potential \cite{CAMARADASILVA2024169549z}. The main appeal of the renormalization process is its ability to relate the problem of boundary conditions to a spontaneous breaking of (asymptotic) conformal symmetry.

\subsection{Renormalization procedure}

Following the construction presented in \cite{CAMARADASILVA2024169549z}, we will adopt an approach that does not require defining boundary conditions at singular points and does not require explicitly choosing a regulator potential. First, the original potential, singular at $x_s$ (from the right or left), is regularized by introducing a cutoff at $R=|x-x_s|$. A new well-defined potential in the interval $0<|x-x_s|<R$ is introduced, i.e.
\begin{eqnarray}
\mathcal V_{R}(x) = \left\{ \begin{array}{ll}
\mathcal{V}(x) & |x-x_s|>R,\\
 & \\
\frac{\lambda(R)}{R^2}f(\frac{|x-x_s|}{R}) & 0<|x-x_s|<R.
\end{array} \right.
\end{eqnarray}
The arbitrary function $f(\frac{|x-x_s|}{R})$, with $f(1)=1$, is non-singular at the origin - usually chosen as a step function, \cite{renor_orig,renor_m_w_c,U_AB,RG_limit_c,renor_Group_Limit_Cycle}, but it is not necessary to specify it. For the renormalization process to provide $R$-independent physical results, we assume that $R$ is a length scale smaller than the others involved in the system, in this case, $1/\alpha$ and $1/k$. Thus:
\begin{eqnarray}
\alpha R\ll1, \ kR\ll1 \ (k\neq0).
\end{eqnarray}
The first condition is necessary for the potential to be approximated by $\mathcal{V}(x)\approx g/(x-x_s)^2$ near the cutoff, and the second condition will be used in the following calculation. Equation (\ref{Sch_x}) in the regularized region, $0<u=|x-x_s|<R$, becomes (the wave function will be denoted as $\psi_{k,<}$)
\begin{eqnarray}
\frac{d^2\psi_{k,<}(u)}{du^2}+\left(\frac{(kR)^2-\lambda(R)f(u/R)}{R^2}\right)\psi_{k,<}(u)=0.
\end{eqnarray}
Neglecting the term $kR$, since $kR\ll1$, we obtain the equation for $k=0$. Therefore, the wave function inside the regularized region does not depend on $k$ and coincides with the $k=0$ case. Thus
\begin{eqnarray}
\frac{\psi'_{k,<}(u)}{\psi_{k,<}(u)}\Bigg|_{u=R^-}=\frac{\psi'_{0,<}(u)}{\psi_{0,<}(u)}\Bigg|_{u=R^{-}}\equiv \mathcal{F}(R),\label{F_R}
\end{eqnarray}
where $\mathcal{F}(R)$ is independent of $k$. By the continuity of the wave function and its first derivative, the result must be the same when taken from outside the cutoff, $\psi'_{k}(u)/\psi_{k}(u)|_{u=R^+}=\mathcal{F}(R)$. That is, we have
\begin{eqnarray}
\left(\frac{d\psi_{k}(u)}{du}-\mathcal{F}(R)\psi_{k}(u)\right)\Bigg|_{u\rightarrow R^+}=0,\label{Robin}
\end{eqnarray}
and the renormalization process effectively consists of imposing a Robin boundary condition at the regularization point $|x-x_s|\rightarrow R^+$, without the need to specify boundary conditions inside the inaccessible region $|x-x_s|<R$. The form of the function $\mathcal{F}(R)$ depends on the value of the coupling.

\subsection{Weak-medium coupling: $-1/4<g<3/4$ ($0<\nu<1$)}\label{sec:beta}
The function $\mathcal{F}(r)$ is easily obtained by substituting the asymptotic form (\ref{x_s}) or (\ref{x_c}) with $|x-x_s|\rightarrow u$, $x_s=0,\pi/2\alpha$, valid for any potential with an inverse-square singularity, for $k=0$
\begin{eqnarray}
R\mathcal{F}(R)=R\frac{\psi_0'(R)}{\psi_0(R)}=\frac{1}{2}+\nu\left(\frac{1+\varepsilon\left(\frac{L_{\nu}}{R}\right)^{2\nu}}{1+\varepsilon\left(\frac{L_{\nu}}{R}\right)^{2\nu}}\right),\label{F_k0}
\end{eqnarray}
where $\varepsilon L_{\nu}^{2\nu}=B_{\nu}(0)/A_{\nu}(0)$, with $L_{\nu}\ge0$ and $\varepsilon=\pm1$ (the relative sign between $B_{\nu}(0)$ and $A_{\nu}(0)$). Each singularity contributes with a length scale $L_{\nu}$. In the case of the TPT potential, we have $L_{\nu_s}$ and $L_{\nu_c}$, which, when finite, spontaneously break the conformal symmetries near the singular points.

Redoing the calculation of equation (\ref{F_k0}) for $k\neq0$, it follows that the parameters $A_{\nu}(k)$ and $B_{\nu}(k)$ in equations (\ref{x_s}) and (\ref{x_c}) are related by
\begin{eqnarray}
\frac{B_{\nu}(k)}{A_{\nu}(k)}=\varepsilon (L_{\nu})^{2\nu}.\label{ren_g_int}
\end{eqnarray}
The asymptotic conformal symmetry is preserved at the singularity in question in two limits: $i)$ $L=0$ ($B_{\nu}(k)=0$), which we will call the UV fixed point; and $UV$; $ii)$ $L\rightarrow\infty$ ($A_{\nu}(k)=0$), which will be called the IR fixed point. Since the TPT potential has two inverse-square singularities, the renormalization process alone implies that if both couplings are in the weak-medium regime (pink square in the center of Figure \ref{fig:esp_g}), then the eigenfunctions and their respective eigenstates should form two-parameter families ($L_{\nu_s},L_{\nu_c}$), one-parameter families $\{(L_{\nu_s},0)$, $(L_{\nu_s},\infty)$, $(0,L_{\nu_c})$, $(\infty,L_{\nu_c})\}$, or no parameters $\{(0,0)$, $(0,\infty)$ , $(\infty,0)$, $(\infty,\infty)\}$. On the other hand, as described in Section \ref{sec:2}, in this case, there is SUSY, and both singularities match UV fixed points, i.e., $L_{\nu_s}=L_{\nu_c}=0$, so there are no parameters to be determined. However, this does not mean that a renormalization in the weak-medium region is irrelevant to the TPT potential. We can have one coupling in this region and the other in the strong-attractive regime (yellow areas in Figures \ref{fig:esp_g} and \ref{fig:esp_g_2}). The limiting case $g=-1/4$ ($\nu=0$) is discussed separately in Appendix \ref{sec:app}.

\subsection{Strong-attractive coupling: $g<-1/4$  ($\nu\rightarrow i|\nu|$)}
In this region, the function $\mathcal{F}(R)$ is determined from the asymptotic form (\ref{x_<}) with $k=0$. Since the potential has its own length scale $1/\alpha$, it can be written as
\begin{eqnarray}
\psi_{\nu,0}(u)\approx |C(0)|x^{1/2}\sin\left(|\nu|\ln(\alpha u)+\theta_{\nu}\right), u=|x-x_s|,\label{x_<_0}
\end{eqnarray}
where $-\pi\le\theta_{\nu}<\pi$ is a constant. With this form, the function $\mathcal{F}(R)$ becomes
\begin{eqnarray}
R\mathcal{F}(R)=\frac{1}{2}+|\nu|\cot\left(|\nu|\ln(\alpha R)+\theta_{\nu}\right).\label{F_k0_2}
\end{eqnarray}
Repeating the calculation for $k\neq0$, along with equation (\ref{x_<}), we find
\begin{eqnarray}
|\nu|\ln\left(k/\alpha\right)+\zeta_{|\nu|}(k)=\theta_{\nu}-n\pi,
\end{eqnarray}
which can also be read as
\begin{eqnarray}
-\frac{1}{\alpha^{2i|\nu|}}\frac{A_{i|\nu|}(k)}{B_{i|\nu|}(k)}\equiv\left(\frac{k}{\alpha}\right)^{2i|\nu|}e^{2i\zeta_{|\nu|}(k)}=e^{i\theta_{\nu}}.\label{ren_g_forte}
\end{eqnarray}
Similarly to the weak-medium coupling, the ratio $B_{i|\nu|(k)}/A_{i|\nu|(k)}$ is once again independent of $k$, but now it is a phase. The renormalization of each singularity with a strong-attractive coupling introduces a phase $\theta_{\nu}$, which generates a one-parameter family for the eigenfunctions given by equation (\ref{sol_G}) and their eigenvalues. Since the choices $A_{i|\nu|}(k)=0$ or $B_{i|\nu|}(k)=0$ are not possible, we are outside the fixed points, and the asymptotic conformal symmetry is spontaneously broken. In the context of linear stability of fields in asymptotically AdS geometries, the strong-attractive coupling region violates the BF bound. In the next section, we will see that when at least one singularity of the TPT potential is strong-attractive, the renormalized spectrum is discrete and well-defined.

%%%%%%%%%%%%%%%%%%%%%%%%%%%%%%%%%%%%%%%%%%%%%%
\section{Renormalized spectrum - $g_s<-1/4$ ($\nu_s\rightarrow i|\nu_s|$)}\label{sec:espectro}

If at least one of the couplings of the TPT potential is of the strong-attractive type, the supersymmetry described in Section \ref{sec:3} is no longer present, and renormalization becomes necessary. Since the couplings are equivalent, we will take the singularity at $x=0$ as a reference, i.e., $g_s<-1/4$ with the condition (\ref{ren_g_forte}) valid for $\nu=\nu_s$. Throughout this section, we will address the three distinct regions - blue, yellow, and green in the lower part of Figure \ref{fig:esp_g}.

\subsection{$g_c\ge3/4$ ($\nu_c>1$)}

We are in the blue region on the lower right side of Figure \ref{fig:esp_g}, and it is necessary to set $B_{\nu_c}(k)=0$ for the eigenfunction to be square-integrable. Thus, equation (\ref{master_2}) is interpreted as follows:
\begin{eqnarray}
0=\frac{-
\frac{\Gamma(\nu_c)\Gamma(1-i|\nu_s|)}{\Gamma\left(\frac{1}{2}-i\frac{|\nu_s|}{2}+\frac{\nu_c}{2}+\frac{k}{2\alpha}\right)\Gamma\left(\frac{1}{2}-i\frac{|\nu_s|}{2}+\frac{\nu_c}{2}-\frac{k}{2\alpha}\right)}e^{-i\theta_{\nu_s}}+\frac{\Gamma(\nu_c)\Gamma(1+i|\nu_s|)}{\Gamma\left(\frac{1}{2}+i\frac{|\nu_s|}{2}+\frac{\nu_c}{2}+\frac{k}{2\alpha}\right)\Gamma\left(\frac{1}{2}+i\frac{|\nu_s|}{2}+\frac{\nu_c}{2}-\frac{k}{2\alpha}\right)}}{-
\frac{\Gamma(-\nu_c)\Gamma(1-i|\nu_s|)}{\Gamma\left(\frac{1}{2}-i\frac{|\nu_s|}{2}-\frac{\nu_c}{2}+\frac{k}{2\alpha}\right)\Gamma\left(\frac{1}{2}-i\frac{|\nu_s|}{2}-\frac{\nu_c}{2}-\frac{k}{2\alpha}\right)}e^{-i\theta_{\nu_s}}+\frac{\Gamma(-\nu_c)\Gamma(1+i|\nu_s|)}{\Gamma\left(\frac{1}{2}+i\frac{|\nu_s|}{2}-\frac{\nu_c}{2}+\frac{k}{2\alpha}\right)\Gamma\left(\frac{1}{2}+i\frac{|\nu_s|}{2}-\frac{\nu_c}{2}-\frac{k}{2\alpha}\right)}}.\label{master_3}
\end{eqnarray}
Solving for the phase $\theta_{\nu_s}$
\begin{eqnarray}
e^{i\theta_{\nu_s}}=\frac{\Gamma(1-i|\nu_s|)}{\Gamma(1+i|\nu_s|)}\frac{\Gamma\left(\frac{1}{2}+i\frac{|\nu_s|}{2}+\frac{\nu_c}{2}+\frac{k}{2\alpha}\right)\Gamma\left(\frac{1}{2}+i\frac{|\nu_s|}{2}+\frac{\nu_c}{2}-\frac{k}{2\alpha}\right)}{\Gamma\left(\frac{1}{2}-i\frac{|\nu_s|}{2}+\frac{\nu_c}{2}+\frac{k}{2\alpha}\right)\Gamma\left(\frac{1}{2}-i\frac{|\nu_s|}{2}+\frac{\nu_c}{2}-\frac{k}{2\alpha}\right)}\equiv e^{{i\Theta_1(k)}}.\label{g_1}
\end{eqnarray}
%%%%%%%%%%%%%%%%%%%%%%%%%%%%%%%%%%%%%%%%%%%%%%%%%%
\begin{figure}[ht] 
\centering
\subfigure[]{
\includegraphics[scale=0.19]{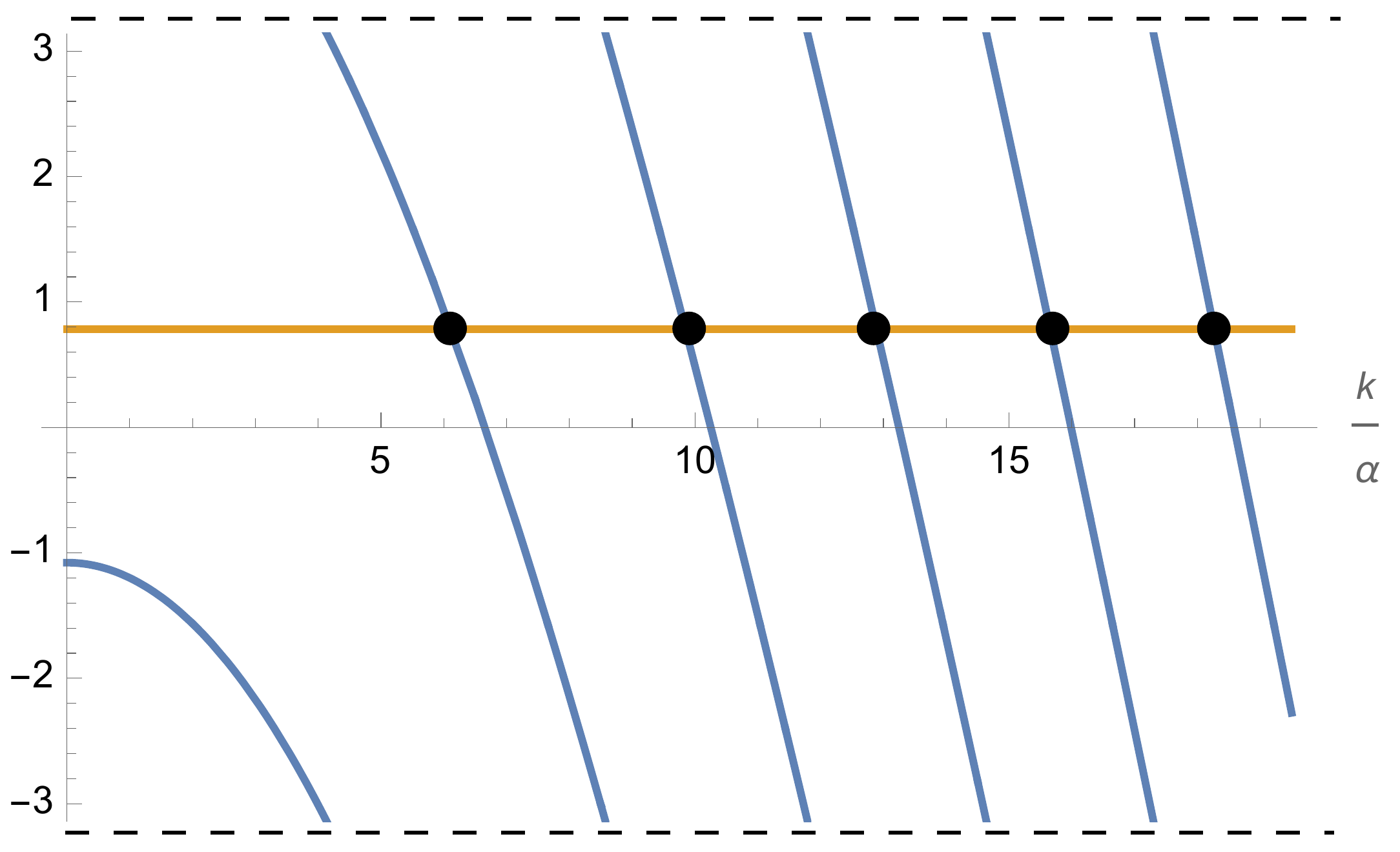}
\label{fig:g_1_k_p}
}
\subfigure[]{
\includegraphics[scale=0.19]{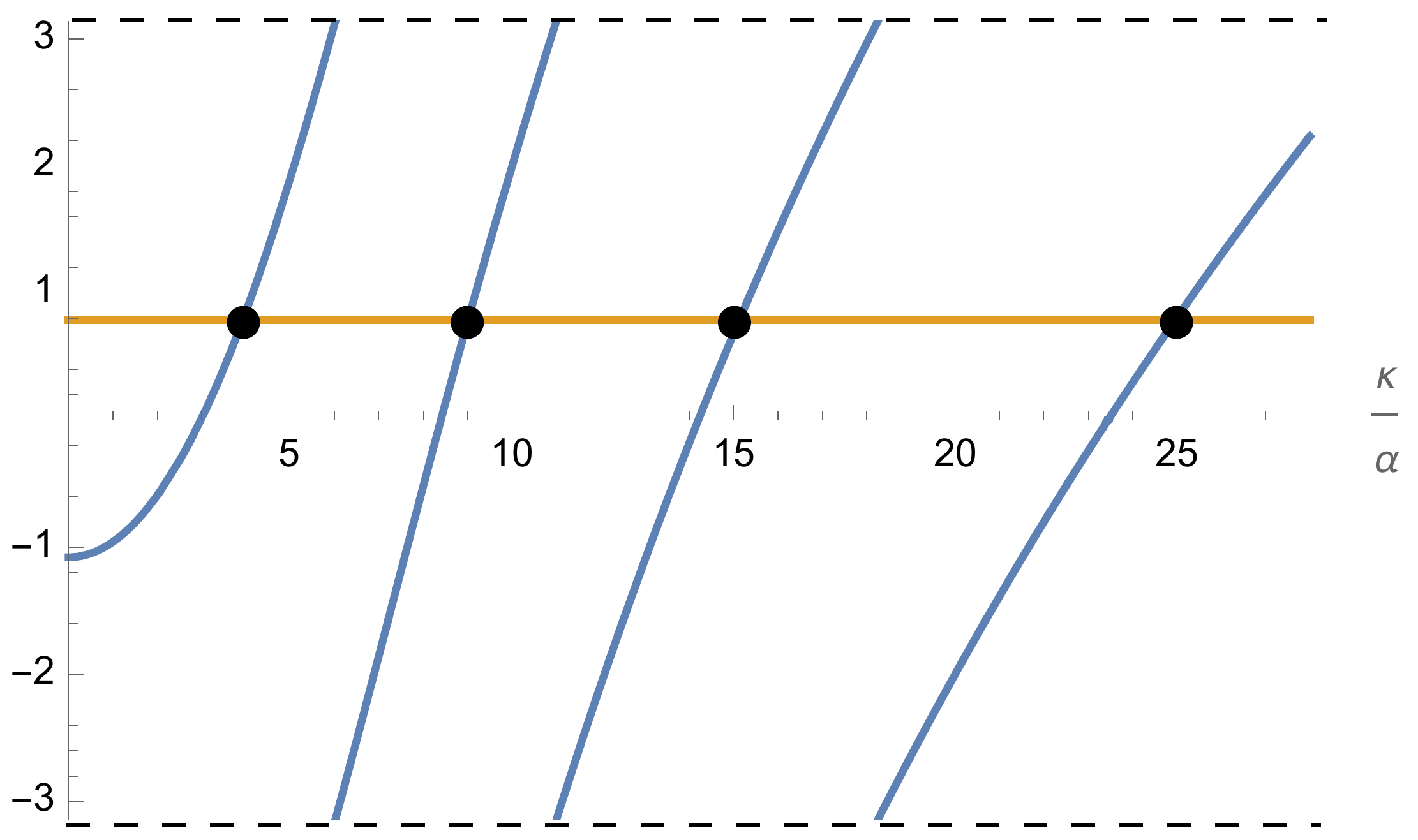}
\label{fig:g_1_k_i}
}
\caption{Curves of $\Theta_1(k)\in[-\pi,\pi)$, equation (\ref{g_1}), for $|\nu_s|=2\pi$ and $\nu_{c}=3.5$. (a) Positive energies ($k>0$); (b) Negative energies ($k\rightarrow i\kappa$). In both cases, the spectrum is the black points where the curves touch the line $\theta_{\nu_s}$, chosen here as $\pi/4$.
}
\label{fig:g_1}
\end{figure}
%%%%%%%%%%%%%%%%%%%%%%%%%%%%%%%%%%%%%%%%%%%%%%%%%%
The result is consistent, as the right-hand side of the equality is a phase, defined as $\Theta_1(k)$ (since the denominator is the complex conjugate of the numerator), even for negative energies, $k\rightarrow i\kappa$. Fixing $\theta_{\nu_s}$, the energy spectrum is determined. In Figures \ref{fig:g_1_k_p} and \ref{fig:g_1_k_i}, we show the graphical solution for the first levels in the case of positive and negative energies, respectively, for the parameter choice $|\nu_s|=2\pi$, $\nu_c=3.5$ and $\theta_{\nu_s}=\pi/4$. In the negative spectrum, there is no accumulation of states near $\kappa=0$ (\emph{Efimov effect}, \cite{renor_m_w_c}), but the deeper states ($\kappa/\alpha\gg1$) tend to obey the \emph{Efimov Rule}
\begin{eqnarray}
\frac{\kappa_{n+1}}{\kappa_n}=e^{\frac{\pi}{|\nu_s|}},
\end{eqnarray}
the exact relation found for $\mathcal{V}(x)=g_s/x^2$, \cite{Bawin:2003dm}. That can be explicitly verified through the values $\kappa_n/\alpha$ presented in Table \ref{tabela}, where the same parameters indicated in Figure \ref{fig:g_1} were used, in particular
\begin{eqnarray}
\frac{\kappa_{7}}{\kappa_{6}}=1.6473, \ \frac{\kappa_{8}}{\kappa_{7}}=1.6482, \ \frac{\kappa_{9}}{\kappa_{8}}=1.6485 \approx e^{\frac{\pi}{|\nu_s|}}=1.6487.
\end{eqnarray}
The interpretation is that deep bound states are near the attractive singularity at $x=0$, so the interaction is effectively $\mathcal{V}(x)\approx g_s/x^2$.

\subsection{$-1/4<g_c<3/4$ ($0<\nu_c<1$)}

The second case corresponds to the yellow region in the lower part of Figure \ref{fig:esp_g} in the coupling space. Without SUSY, there is no criterion to fix the scale $L_{\nu_c}$ to a specific value. Equation (\ref{master_2}) with the renormalized boundary condition (\ref{ren_g_int}) can be solved for the scale $L_{\nu_c}$
%%%%%%%%%%%%%%%%%%%%%%%%%%%%%%%%%%%%%%%%%%%%%%%%%%%%%%%%
\begin{figure}[ht] 
\centering
\subfigure[]{
\includegraphics[scale=0.18]{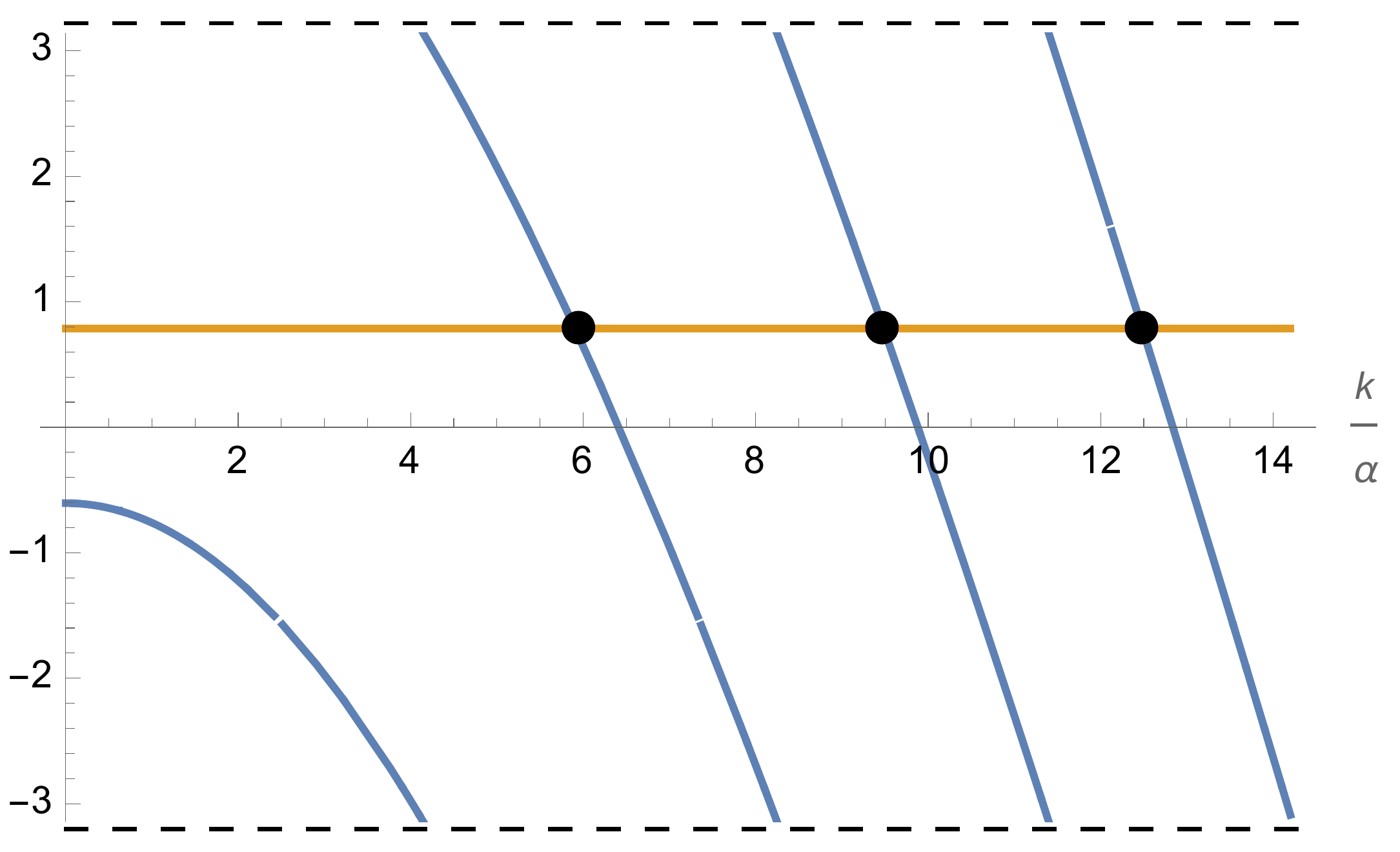}
\label{fig:g_2_k_p}
}
\subfigure[]{
\includegraphics[scale=0.18]{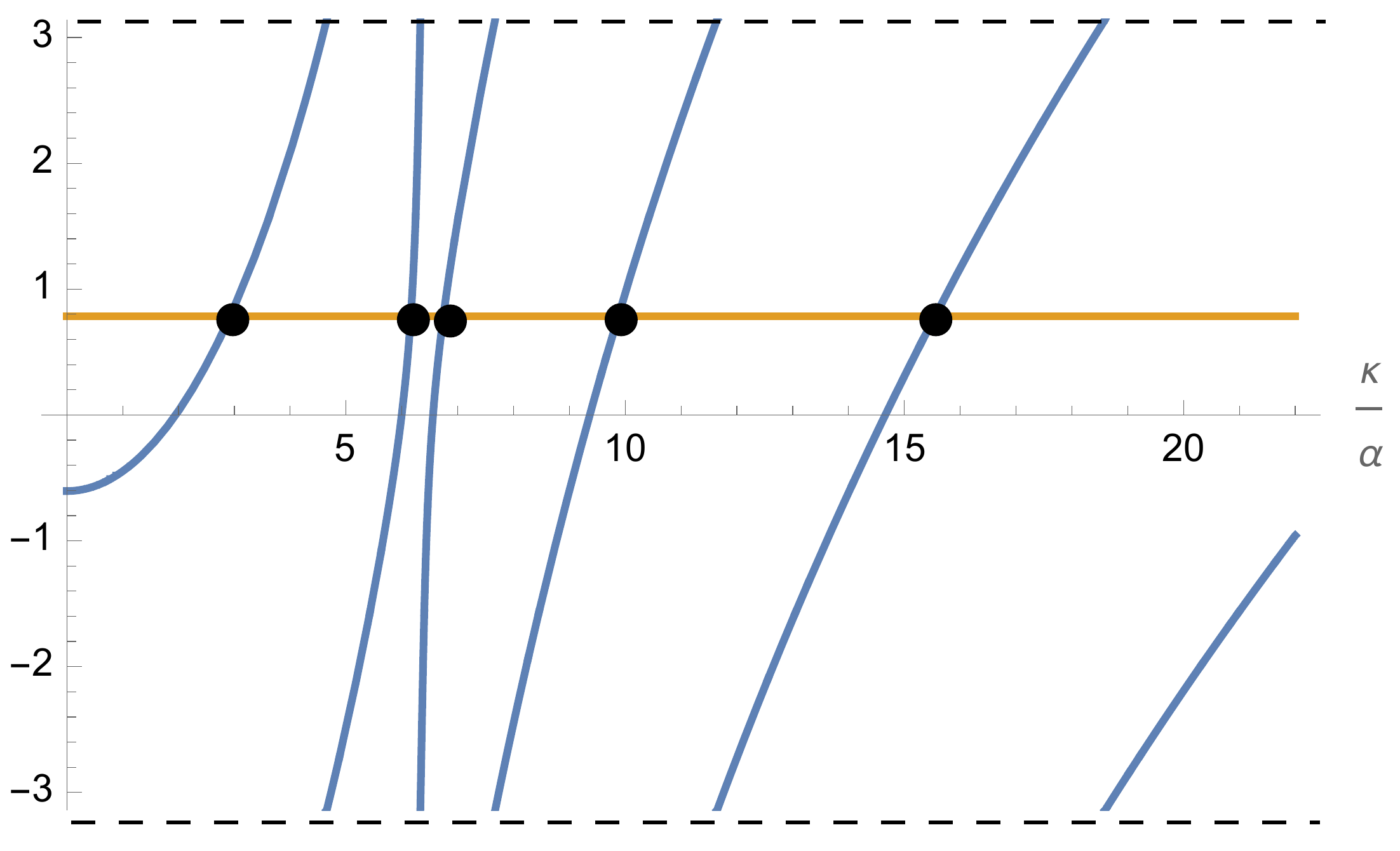}
\label{fig:g_2_k_i}
}
\caption{Curve $\Theta_2(k)$, equation (\ref{master_4_2}), for $|\nu_s|=2\pi$, $\nu_{c}=0.9$, and $\varepsilon(\alpha L_{\nu_c})^{2\nu_c}=0.6$. (a) Positive energies ($k>0$); (b) Negative energies ($k\rightarrow i\kappa$). In both cases, the spectrum is the points where the curves touch the line $\theta_{\nu_s}$, chosen as $\pi/4$.
}
\label{fig:g_2}
\end{figure}
%%%%%%%%%%%%%%%%%%%%%%%%%%%%%%%%%%%%%%%%%%%%%%%%%%%%%%%%
\begin{eqnarray}
\!\!\!\!\!&\varepsilon\left(\alpha L_{\nu_c}\right)^{2\nu_c}=\frac{-
\frac{\Gamma(\nu_c)\Gamma(1-i|\nu_s|)}{\Gamma\left(\frac{1}{2}-i\frac{|\nu_s|}{2}+\frac{\nu_c}{2}+\frac{k}{2\alpha}\right)\Gamma\left(\frac{1}{2}-i\frac{|\nu_s|}{2}+\frac{\nu_c}{2}-\frac{k}{2\alpha}\right)}e^{-i\theta_{\nu_s}}+\frac{\Gamma(\nu_c)\Gamma(1+i|\nu_s|)}{\Gamma\left(\frac{1}{2}+i\frac{|\nu_s|}{2}+\frac{\nu_c}{2}+\frac{k}{2\alpha}\right)\Gamma\left(\frac{1}{2}+i\frac{|\nu_s|}{2}+\frac{\nu_c}{2}-\frac{k}{2\alpha}\right)}}{-
\frac{\Gamma(-\nu_c)\Gamma(1-i|\nu_s|)}{\Gamma\left(\frac{1}{2}-i\frac{|\nu_s|}{2}-\frac{\nu_c}{2}+\frac{k}{2\alpha}\right)\Gamma\left(\frac{1}{2}-i\frac{|\nu_s|}{2}-\frac{\nu_c}{2}-\frac{k}{2\alpha}\right)}e^{-i\theta_{\nu_s}}+\frac{\Gamma(-\nu_c)\Gamma(1+i|\nu_s|)}{\Gamma\left(\frac{1}{2}+i\frac{|\nu_s|}{2}-\frac{\nu_c}{2}+\frac{k}{2\alpha}\right)\Gamma\left(\frac{1}{2}+i\frac{|\nu_s|}{2}-\frac{\nu_c}{2}-\frac{k}{2\alpha}\right)}}.\label{master_4}
\end{eqnarray}
Inverting for the phase $\theta_{\nu_s}$
\begin{eqnarray}
\!\!\!\!\!&e^{-i\theta_{\nu_s}}=\frac{\frac{\varepsilon\Gamma(-\nu_c)\Gamma(1+i|\nu_s|)}{\Gamma\left(\frac{1}{2}+i\frac{|\nu_s|}{2}-\frac{\nu_c}{2}+\frac{k}{2\alpha}\right)\Gamma\left(\frac{1}{2}+i\frac{|\nu_s|}{2}-\frac{\nu_c}{2}-\frac{k}{2\alpha}\right)}(\alpha L_{\nu_c})^{2\nu_c}-\frac{\Gamma(\nu_c)\Gamma(1+i|\nu_s|)}{\Gamma\left(\frac{1}{2}+i\frac{|\nu_s|}{2}+\frac{\nu_c}{2}+\frac{k}{2\alpha}\right)\Gamma\left(\frac{1}{2}+i\frac{|\nu_s|}{2}+\frac{\nu_c}{2}-\frac{k}{2\alpha}\right)}}{
\frac{\varepsilon\Gamma(-\nu_c)\Gamma(1-i|\nu_s|)}{\Gamma\left(\frac{1}{2}-i\frac{|\nu_s|}{2}-\frac{\nu_c}{2}+\frac{k}{2\alpha}\right)\Gamma\left(\frac{1}{2}-i\frac{|\nu_s|}{2}-\frac{\nu_c}{2}-\frac{k}{2\alpha}\right)}(\alpha L_{\nu_c})^{2\nu_c}-
\frac{\Gamma(\nu_c)\Gamma(1-i|\nu_s|)}{\Gamma\left(\frac{1}{2}-i\frac{|\nu_s|}{2}+\frac{\nu_c}{2}+\frac{k}{2\alpha}\right)\Gamma\left(\frac{1}{2}-i\frac{|\nu_s|}{2}+\frac{\nu_c}{2}-\frac{k}{2\alpha}\right)}}.\label{master_4_2}
\end{eqnarray}
By direct inspection, one can verify that the right-hand side of equation (\ref{master_4}) is a real quantity (it is its conjugate) even when $k\rightarrow i\kappa$, corresponding to negative energy. Similarly, the right-hand side of equation (\ref{master_4_2}) is a phase, which we will call $e^{-i\Theta_2(k)}$. Both equations completely describe the energy spectrum in terms of the pair of numbers $\{\varepsilon(\alpha L_{\nu_c})^{2\nu_c},\theta_{\nu_s}\}$. For illustration, Figure \ref{fig:g_2} shows the first bound states obtained graphically. It displays the curve $\Theta_2(k)$ in blue for $|\nu_s|=2\pi$, $\nu_{c}=0.9$ and $\varepsilon(\alpha L_{\nu_c})^{2\nu_c}=0.6$. In Figure \ref{fig:g_2_k_p}, the energy is positive ($k>0$), and in Figure \ref{fig:g_2_k_i}, the energy is negative ($k\rightarrow i\kappa$, $\kappa>0$). The bound states are the black points where $\Theta_2(k)$ touches the line $\theta_{\nu_s}$, chosen as $\pi/4$. For negative energies, this case is qualitatively similar to the previous one: for small energies (in absolute value), the wave function interacts with the entire potential, and there is no creation of the \emph{Efimov effect} -accumulation of bound states in the low-energy (in absolute value) region. On the other hand, for deeper states, the wave function tends to be confined near the singularity at $x=0$, and effectively, we have a potential $\mathcal{V}(x)\approx g/x^2$. This can be easily verified by taking the ratio $\kappa_{n+1}/\kappa_{n}$ for some results presented in Table \ref{tabela} (the same values used in the caption of Figure \ref{fig:g_2} were used), e.g.
\begin{eqnarray}
\frac{\kappa_9}{\kappa_8}= 1.6473, \ \frac{\kappa_{10}}{\kappa_9}=1.6481, \ \frac{\kappa_{11}}{\kappa_{10}}= 1.6485\approx e^{\frac{\pi}{|\nu_s|}}=1.6487,
\end{eqnarray}
which tends to increasingly approach the \emph{Efimov Rule}, characteristic of the inverse-square potential.
%%%%%%%%%%%%%%%%%%%%%%%%%%%%%%%%
\subsection{$g_c<-1/4$ ($\nu_c\rightarrow i|\nu_c|$)}

That is the last region to be analyzed, corresponding to the green part on the lower left side of Figure \ref{fig:esp_g}. Here, in addition to $g_s$, the coupling $g_c$ is also in the strong-attractive regime ($\nu_c\rightarrow i|\nu_c|$). Equation (\ref{master_2}) determines the energy spectrum through
\begin{eqnarray}
\!\!\!\!\!\!\!&e^{-i\theta_{\nu_c}}=\frac{
\frac{\Gamma(i|\nu_c|)\Gamma(1-i|\nu_s|)}{\Gamma\left(\frac{1}{2}-i\frac{|\nu_s|}{2}+i\frac{|\nu_c|}{2}+\frac{k}{2\alpha}\right)\Gamma\left(\frac{1}{2}-i\frac{|\nu_s|}{2}+i\frac{|\nu_c|}{2}-\frac{k}{2\alpha}\right)}e^{-i\theta_{\nu_s}}-\frac{\Gamma(i|\nu_c|)\Gamma(1+i|\nu_s|)}{\Gamma\left(\frac{1}{2}+i\frac{|\nu_s|}{2}+i\frac{|\nu_c|}{2}+\frac{k}{2\alpha}\right)\Gamma\left(\frac{1}{2}+i\frac{|\nu_s|}{2}+i\frac{|\nu_c|}{2}-\frac{k}{2\alpha}\right)}}{-
\frac{\Gamma(-i|\nu_c|)\Gamma(1-i|\nu_s|)}{\Gamma\left(\frac{1}{2}-i\frac{|\nu_s|}{2}-i\frac{|\nu_c|}{2}+\frac{k}{2\alpha}\right)\Gamma\left(\frac{1}{2}-i\frac{|\nu_s|}{2}-i\frac{|\nu_c|}{2}-\frac{k}{2\alpha}\right)}e^{-i\theta_{\nu_s}}+\frac{\Gamma(-i|\nu_c|)\Gamma(1+i|\nu_s|)}{\Gamma\left(\frac{1}{2}+i\frac{|\nu_s|}{2}-i\frac{|\nu_c|}{2}+\frac{k}{2\alpha}\right)\Gamma\left(\frac{1}{2}+i\frac{|\nu_s|}{2}-i\frac{|\nu_c|}{2}-\frac{k}{2\alpha}\right)}}.\label{master_5}
\end{eqnarray}
Since both parameters are phases, equation (\ref{master_5}), or its inverted form in favor of $\theta_{\nu_s}$, does not reflect the fact that $\theta_{\nu_s}$ and $\theta_{\nu_c}$ are treated equally. To make the result more symmetric, we will express it in terms of the new quantities $\{\bar\theta,\Delta\theta\}\equiv\{\frac{1}{2}\left(\theta_{\nu_c}+\theta_{\nu_s}\right),\frac{1}{2}\left(\theta_{\nu_c}-\theta_{\nu_s}\right)\}$. The first step is to rewrite equation (\ref{master_5}) using the notation introduced in equation (\ref{M_}), i.e.:
\begin{eqnarray}
&&e^{-i\theta_{\nu_c}}=\frac{\mathcal{M}_1(k)e^{-i\theta_{\nu_s}}-\mathcal{M}_2(k)}{-\mathcal{M}_2^*(k)e^{-i\theta_{\nu_s}}+\mathcal{M}_1^*(k)}, \ |\mathcal{M}_1(k)|^2-|\mathcal{M}_2(k)|^2=-\frac{|\nu_s|}{|\nu_c|},\label{master_5_new}
\end{eqnarray}
where it was used that $\mathcal{M}_4(k)=\mathcal{M}_1^*(k)$ and $\mathcal{M}_3(k)=\mathcal{M}_2^*(k)$, regardless of whether $k/\alpha$ is a real or purely imaginary number. Manipulating Equation (\ref{master_5_new}), we easily get
\begin{eqnarray}
\mathcal{M}_2^*(k)e^{-2i\bar\theta}+\left(\mathcal{M}_1(k)e^{i\Delta\theta}-\mathcal{M}^*_1(k)e^{-i\Delta\theta}\right)e^{-i\bar\theta}-\mathcal{M}_2(k)=0.
\end{eqnarray}
The term in parentheses in the equality above is given by $2i\Im m\left(\mathcal{M}_1(k)e^{i\Delta\theta}\right)$. Solving the quadratic equation for $e^{-i\bar\theta}$,
\begin{eqnarray}
e^{-i\bar\theta}=-i\frac{\Im m\left(\mathcal{M}_1(k)e^{i\Delta\theta}\right)}{\mathcal{M}_2^*(k)}+\frac{1}{\mathcal{M}_2^*(k)}\sqrt{\left|\mathcal{M}_2(k)\right|^2-\left(\Im m\left(\mathcal{M}_1(k)e^{i\Delta\theta}\right)\right)^2}.\label{int}
\end{eqnarray}
Finally, writing $\mathcal{M}_2(k)=|\mathcal{M}_2(k)|e^{i\zeta_2(k)}$ and defining
\begin{eqnarray}
\sin\Sigma(k;\Delta \theta)\equiv \frac{\Im m\left(\mathcal{M}_1(k)e^{i\Delta\theta}\right)}{|\mathcal{M}_2(k)|},
\end{eqnarray}
a quantity always less than unity since $|\mathcal{M}_1(k)|<|\mathcal{M}_2(k)|$ due to the constraint in equation (\ref{master_5_new}). With these definitions, we have the following form for equation (\ref{int})
\begin{eqnarray}
&&e^{-i\bar\theta}=e^{i\zeta_2(k)}\left(-i\sin\Sigma(k;\Delta \theta)+\cos\Sigma(k;\Delta \theta)\right)=e^{-i\left(\Sigma(k;\Delta \theta)-\zeta_2(k)\right)},\nonumber\\
&&\cos\bar\theta-i\sin\bar\theta=\cos\left(\Sigma(k;\Delta \theta)-\zeta_2(k)\right)-i\sin\left(\Sigma(k;\Delta \theta)-\zeta_2(k)\right).\nonumber
\end{eqnarray}
The real and imaginary parts imply two equalities, and their ratio yields
\begin{eqnarray}
\tan\bar\theta=\tan\left(\Sigma(k;\Delta\theta)-\zeta_2(k)\right),\label{atrat_final}
\end{eqnarray}
%%%%%%%%%%%%%%%%%%%%%%%%%%%%%%%%%%%%%%%%%%%%%%%%
\begin{figure}[ht] 
\centering
\subfigure[]{
\includegraphics[scale=0.19]{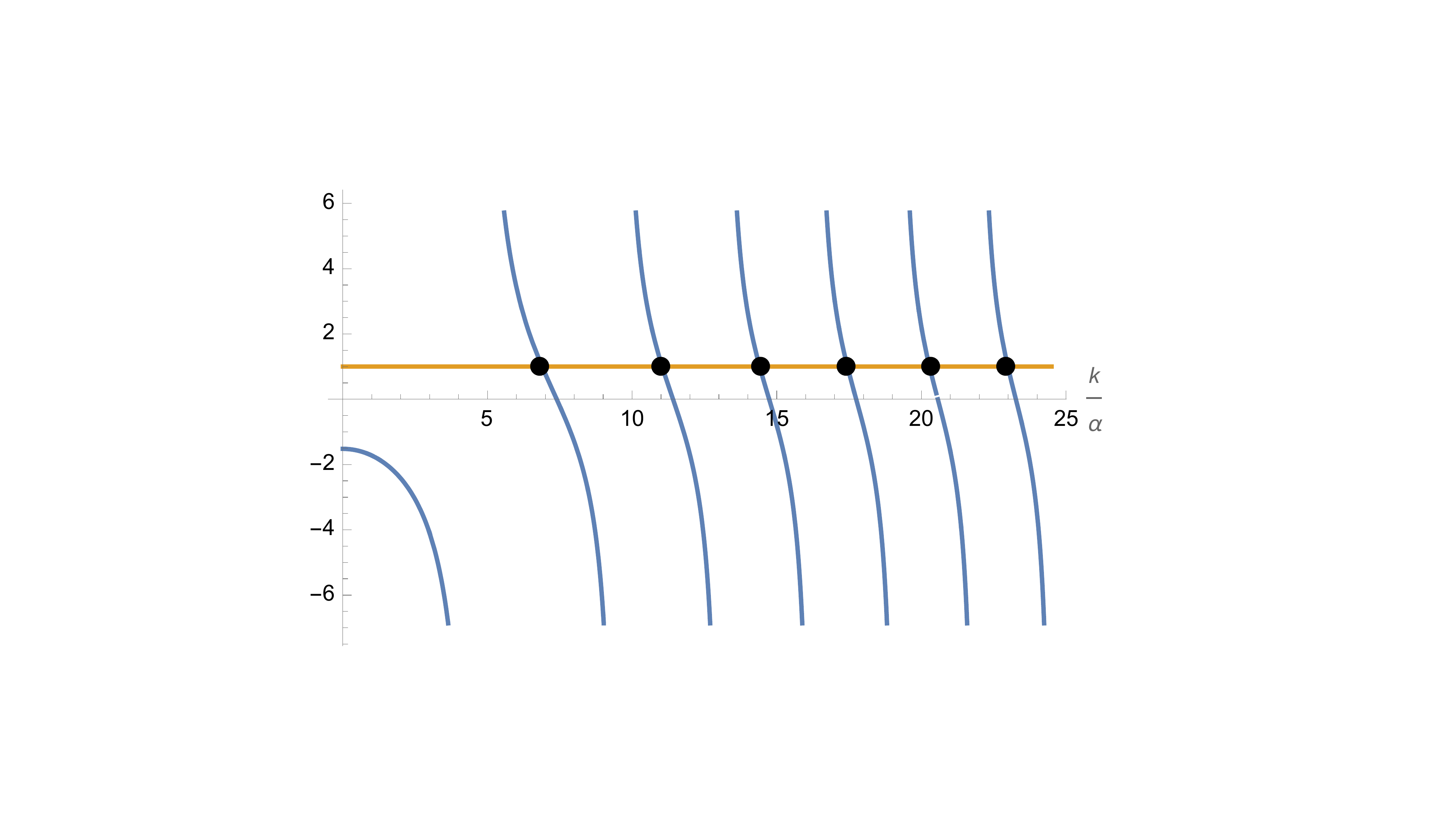}
\label{fig:_forte_1}
}
\subfigure[]{
\includegraphics[scale=0.18]{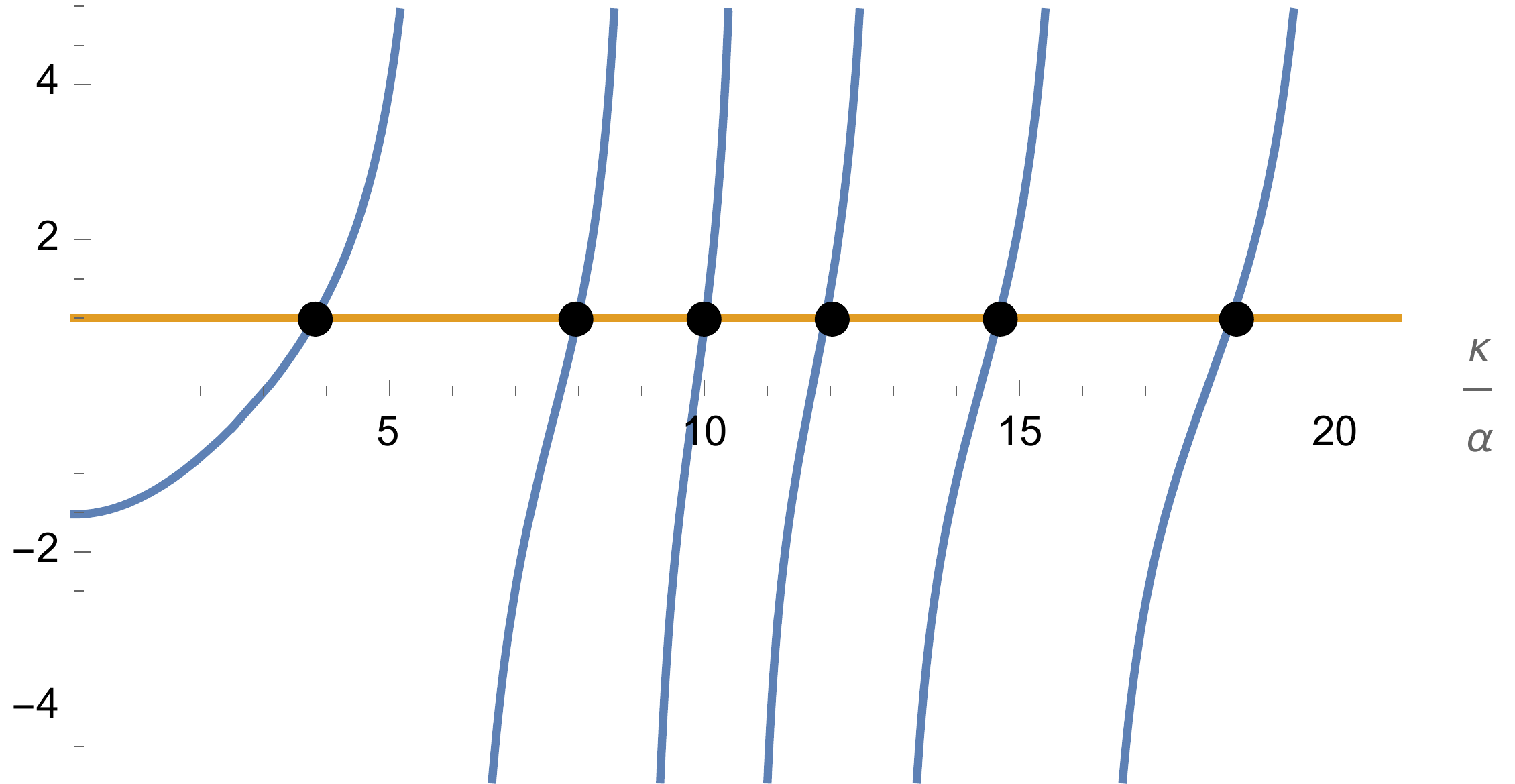}
\label{fig:_forte_2}
}
\caption{The blue curves are the ratio $\tan\left(\Sigma(k;\Delta\theta)-\zeta_2(k)\right)/\tan\left((\theta_{\nu_s}+\theta_{\nu_c})/2\right)$ from equation (\ref{atrat_final}). The bound states are the black points where the curves touch the beige unit line. (a) Positive energies ($k>0$); (b) Negative energies ($k\rightarrow i\kappa$). In both cases, the parameters chosen were $|\nu_s|=2\pi$, $|\nu_c|=\pi$, $\theta_{\nu_s}=\frac{\pi}{3}$, and $\theta_{\nu_c}=-\frac{\pi}{8}$.
}
\label{fig:_forte}
\end{figure}
%%%%%%%%%%%%%%%%%%%%%%%%%%%%%%%%%%%%%%%%%%%%%%%%
the energy spectrum in terms of a transcendental function. In Figures \ref{fig:_forte_1} and \ref{fig:_forte_2}, we have, respectively, the graphical determination of the first positive and negative energy modes for the case $\{|\nu_s|=2\pi,|\nu_c|=\pi;\theta_{\nu_s}=\frac{\pi}{3},\theta_{\nu_c}=-\frac{\pi}{8}\}$. Again, there is no accumulation of bound states with small negative energies, but for large negative energy modes (in absolute value), we have a novelty. Through numerical analysis, we discovered a competition between the singularities, with the \emph{Efimov Rule} realized for both. On the other hand, this occurs with an unexpected factor of $2$ multiplying $|\nu|$, resulting in ratios approaching both $\kappa_{n+1}/\kappa_n\approx e^{\frac{\pi}{2|\nu_c|}}$ and $\kappa_{n+1}/\kappa_n\approx e^{\frac{\pi}{2|\nu_s|}}$, with the realization occurring first for the singularity with the larger $|\nu|$ (smaller $e^{\frac{\pi}{2|\nu|}}$). To illustrate, we take the ratios of some of the values provided in Table \ref{tabela} (whose parameter values are the same as those written in the text and also used in Figure \ref{fig:_forte})
\begin{eqnarray}
&&\frac{\kappa_{10}}{\kappa_{9}}=1.2825, \ \frac{\kappa_{11}}{\kappa_{10}}=1.2831,\ \frac{\kappa_{12}}{\kappa_{11}}=1.2835\approx e^{\frac{\pi}{2|\nu_s|}}=1.2840,\\
&&\frac{\kappa_{14}}{\kappa_{13}}=1.6481\approx e^{\frac{\pi}{2|\nu_c|}}=1.6487.
\end{eqnarray}
On the other hand, these simple rules do not hold in all cases. For example, $\kappa_{13}/\kappa_{12}\approx 1.4523$ and $\kappa_{15}/\kappa_{14}\approx 1.4571$, which we suspect converge slowly to $\frac{1}{2}\left(e^{\frac{\pi}{2|\nu_c|}}+e^{\frac{\pi}{2|\nu_s|}}\right)\approx1.4664$. The interpretation of the results is straightforward: the negative energy eigenstates first tend to nest in the more singular side of the potential. Penetrating further, we eventually have states trapped near the other singularity, and finally, there are states distributed around both singularities. Unfortunately, we didn't determine a pattern for how the different rules alternate.

To conclude, a pertinent comment is that renormalization predicts infinite bound states in the negative energy region in all three cases. However, this conclusion cannot be taken to its ultimate consequences, as the renormalization process ceases to be valid when $\kappa\sim 1/R$.
%%%%%%%%%%%%%%%%%%%%%%%%%%%%%%%%%%%%%%%%%%%%%%%%%%%%%%%%%%%%%%%%%%%%%%%%%%
\begin{center}
\begin{tabular}{|c|c|c|c|}
\hline
 $g_c$   & strongly repulsive & weak medium & strongly attractive \\
 \hline 
 $\frac{\kappa_1}{\alpha}$ & 3.896 & 2.925 & 3.826\\
 \hline 
 $\frac{\kappa_2}{\alpha}$ & 9.898 & 6.159 & 7.990\\
\hline 
 $\frac{\kappa_3}{\alpha}$ & 15.559 & 6.743 & 10.015\\
 \hline 
 $\frac{\kappa_4}{\alpha}$ & 25.200 & 9.875 & 11.921\\
 \hline 
 $\frac{\kappa_5}{\alpha}$ & 41.275 & 15.548 & 14.666\\
 \hline 
 $\frac{\kappa_6}{\alpha}$ & 67.886 & 25.193 & 18.365\\
 \hline 
 $\frac{\kappa_7}{\alpha}$ & 111.825 & 41.271 & 26.523\\
 \hline 
 $\frac{\kappa_8}{\alpha}$ & 184.307 & 67.883 & 42.862\\
 \hline 
 $\frac{\kappa_9}{\alpha}$ & 303.834 & 111.823 & 61.913\\
 \hline 
 $\frac{\kappa_{10}}{\alpha}$ & - & 184.306 & 79.402\\
 \hline 
 $\frac{\kappa_{11}}{\alpha}$ & - & 303.834 & 101.879\\
 \hline 
 $\frac{\kappa_{12}}{\alpha}$ & - & - & 130.758\\
 \hline 
 $\frac{\kappa_{13}}{\alpha}$ & - & - & 189.933\\
 \hline 
 $\frac{\kappa_{14}}{\alpha}$ & - & - & 313.037\\
 \hline
 $\frac{\kappa_{15}}{\alpha}$ & - & - & 456.103\\
 \hline
\end{tabular}
\captionof{table}{Numerical values for negative energy states that satisfy equations (\ref{g_1}) (strongly repulsive $g_{c}$), (\ref{master_4_2}) (weak medium $g_c$), and (\ref{atrat_final}) (strongly attractive $g_c$). In all cases, the parameters are the same as those used in the text and also in Figures \ref{fig:g_1}, \ref{fig:g_2}, and \ref{fig:_forte}, respectively. The values in the first two columns tend to agree because we used the same $\theta_{\nu_s}$ and $|\nu_c|$, the only relevant parameters for the deepest levels, where the states are near the singularity at $x=0$.}
\label{tabela}
\end{center}

\section{Symmetric double-well trigonometric P\"oschl-Teller potential}\label{sec:duplo}

The renormalization of the TPT model introduces families of up to two parameters of energy eigenstates. Since the renormalized Hamiltonian is self-adjoint, each family of eigenstates forms an orthogonal basis of the Hilbert space in the interval $0<x<\frac{\pi}{2\alpha}$. Extending the potential to the interval $-\frac{\pi}{2\alpha}<x<\frac{\pi}{2\alpha}$, we have a symmetric well, so its eigenstates must have well-defined parity around the origin, \cite{Landau_3}. What distinguishes the TPT well from a usual symmetric well is the presence of a singularity at $x=0$, making the construction of the energy spectrum non-trivial. Simply extending an already determined family of eigenfunctions with even and odd parity would only create a duplication with a degenerate spectrum that does not form a basis of the extended Hilbert space. To have a basis in the new interval, we need to renormalize each side of the singularity at the origin independently, i.e., there is a renormalization for the interval $-\frac{\pi}{2\alpha}<x<0^-$ and another for $0^+<x<\frac{\pi}{2\alpha}$. On the other hand, since the parity of the eigenstates is well-defined, one side is duplicated in an even manner and the other in an odd manner. The even modes must satisfy Neumann boundary conditions, $d\psi_{k;\nu_s,\nu_c}(0)/dx=0$, while the odd modes obey Dirichlet conditions, $\psi_{k;\nu_s,\nu_c}(0)=0$. Unlike the original interval, in the double well, there is no SUSY to fix a priori the scales $L_{\nu_s}^{\pm}$ as zero in the weak-medium region, and coincidentally, only in the weak-medium region of the coupling $g_s$ (central brown, pink, and yellow band in Figure \ref{fig:esp_g}) is it possible to meet all the requirements for the construction of the double-well eigenstates in a non-trivial way. This occurs because it contains the particular value $\nu_s=1/2$ ($g_s=0$), where the condition (\ref{F_k0}) in the limit $R\rightarrow0$ (well-defined in this case) leads to the Robin boundary condition
\begin{eqnarray}
\frac{1}{\varepsilon L^{\pm}_{1/2}}\psi_{k;1/2,\nu_c}(0^{\pm})+\psi'_{k;1/2,\nu_c}(0^{\pm})=0,
\end{eqnarray}
where $L^{\pm}_{1/2}=0$ corresponds to the Dirichlet condition and $L^{\pm}_{1/2}\rightarrow\infty$ corresponds to the Neumann condition. Interpreting $\nu_c$ as a parameter that changes adiabatically within each regime, then for other values of $\nu_s$ in the weak-medium region ($0<\nu_s<1$), where the limit $R\rightarrow0$ is not possible, the solution in the region $|x|<R$ must be constructed so as to maintain the same boundary condition as for $\nu_s=1/2$, i.e., Dirichlet for IR$_{\nu_s}$ and Neumann for UV$_{\nu_s}$ - obviously, the previous results are not altered as renormalization is independent of the boundary condition at the origin. However, now we have a clear criterion (the parity of the eigenstates) that uniquely determines this boundary condition. One of the scales (say $L_{\nu_s}^+$) is necessarily zero, and the eigenstates defined in $0<x<\frac{\pi}{2\alpha}$ are extended oddly to the other interval, while the other scale (say $L_{\nu}^-$), is taken as infinite, and the eigenstates given in $-\frac{\pi}{2\alpha}<x<0$ are extended in an even manner. Note that the argument is compatible with the fact that if $1/2<\nu_s<1$, at the UV$_{\nu_s}$ fixed point, the eigenfunctions grow near the origin, and the even extension is natural while the odd extension is not possible. On the other hand, at the IR$_{\nu_s}$ fixed point, the eigenfunctions always tend to zero, which is a necessary condition for the odd extension. When $0<\nu_s<1/2$, the eigenfunctions tend to zero in both cases and, at the UV$_{\nu_s}$ fixed point, it is necessary to impose the Neumann condition in the non-physical region (which does not alter the procedure presented). The renormalizations at the points $\pm\frac{\pi}{2\alpha}$ can, a priori, introduce different scales, $L_{\nu_c}^{\pm}$, for the even and odd states, but we see no physical reason for this and will maintain the same scale at both points, $L_{\nu_c}^+=L_{\nu_c}^-\equiv L_{\nu_c}$ - see Subsection \ref{sec:part} for an exception.

\subsection{$g_c>3/4$ ($\nu_c>1$)}
\begin{figure}[ht] 
\centering
\subfigure[]{
\includegraphics[scale=0.5]{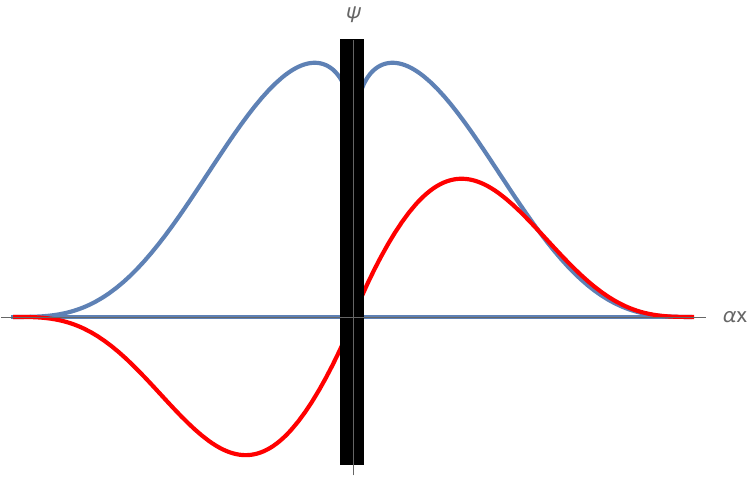}
\label{fig:psi_1_a}
}
\subfigure[]{
\includegraphics[scale=0.5]{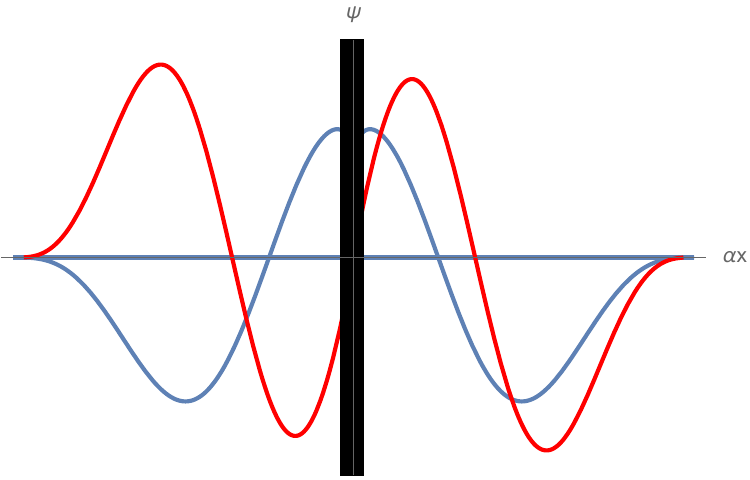}
\label{fig:psi_1_b}
}
\caption{Unnormalized energy eigenstates of the double well associated with $\nu_s=0.4$ and $\nu_c=2.5$ ($g_c>3/4$): (a) Ground state in blue and first excited state in red. (b) Second excited state in blue and third excited state in red. The black bands represent the region $|x|<R$.
}
\label{fig:_duplo_1}
\end{figure}

We are in the brown region on the right side of Figure \ref{fig:esp_g}, where the IR$_{\nu_c}$ fixed point is selected, i.e., $B_{\nu_c}(k)=0$. From equation (\ref{master}), the spectra of the even and odd eigenfunctions are given, respectively, by
\begin{eqnarray}
&&\frac{k_n^{(even)}}{\alpha}=2n+1-\nu_s+\nu_c, \ n\in\mathbb{Z}_{\ge0}, \ (A_{\nu_s}(k)=0)\label{k_UV_IR}\\
&&\frac{k_n^{(odd)}}{\alpha}=2n+1+\nu_s+\nu_c, \ n\in\mathbb{Z}_{\ge0}, \ (B_{\nu_s}(k)=0).\label{k_IR_IR}
\end{eqnarray}
By direct inspection, it is easy to see that the spectrum alternates between even and odd modes following the order $k_{0}^{(even)}<k_{0}^{(odd)}<k_{1}^{(even)}<k_{1}^{(odd)}<\ldots$. In Figures \ref{fig:psi_1_a} and \ref{fig:psi_1_b}, we have the first four energy eigenstates for $\nu_s=0.4$ and $\nu_c=2.5$. They show that, even with the singularity, the \emph{Node Theorem} remains valid, \cite{Moriconi_Nodes}.
%%%%%%%%%%%%%%%%%%%%%%%%%%%%%%%%%%%%%%%%%%%%%%%%
%%%%%%%%%%%%%%%%%%%%%%%%%%%%%%%%%%%%%%%%%%%%%%%%t 
\subsection{$-1/4<g_c<3/4$ ($0<\nu_c<1$)}
\begin{figure}[ht] 
\centering
\subfigure[]{
\includegraphics[scale=0.15]{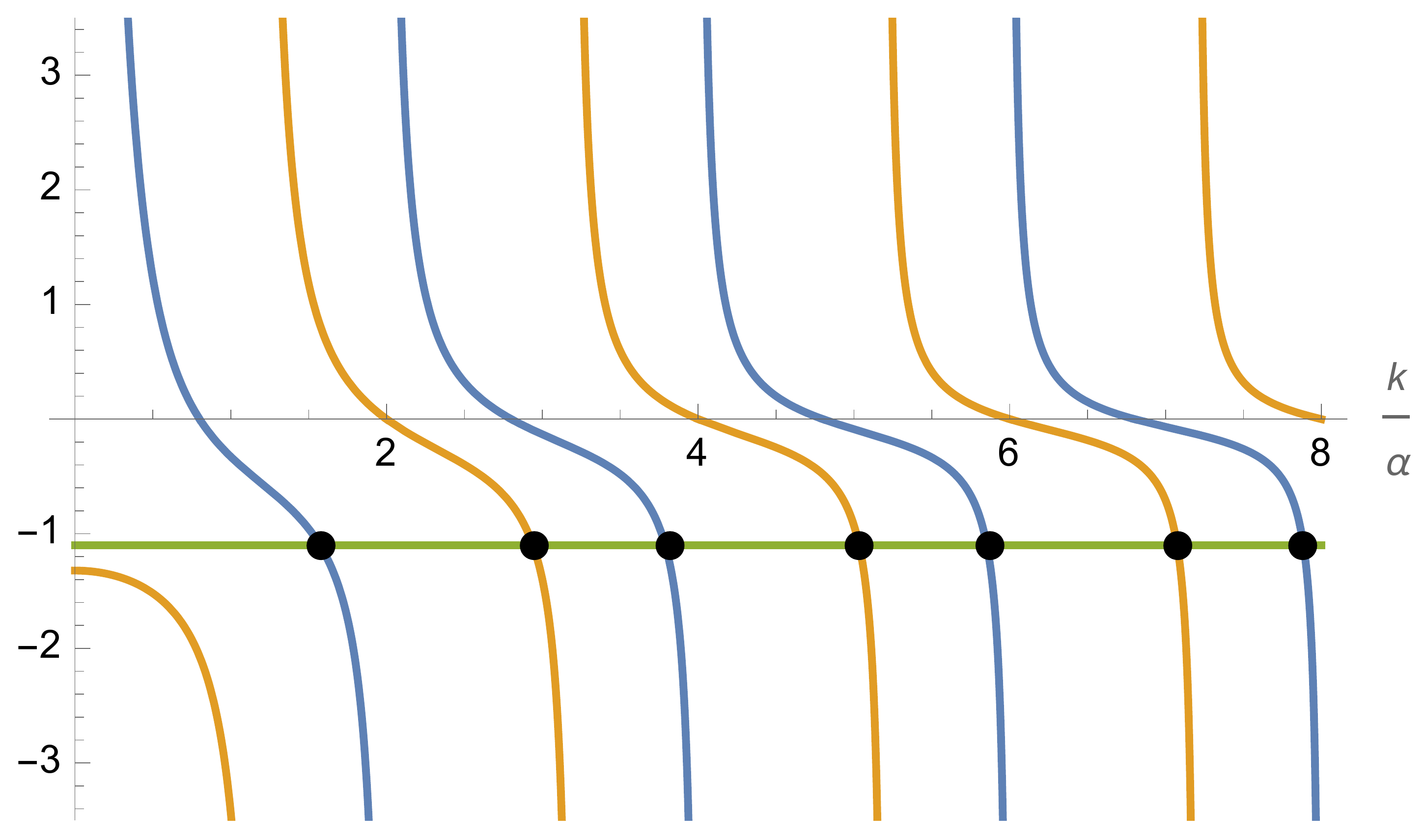}
\label{fig:poco_1}
}
\subfigure[]{
\includegraphics[scale=0.15]{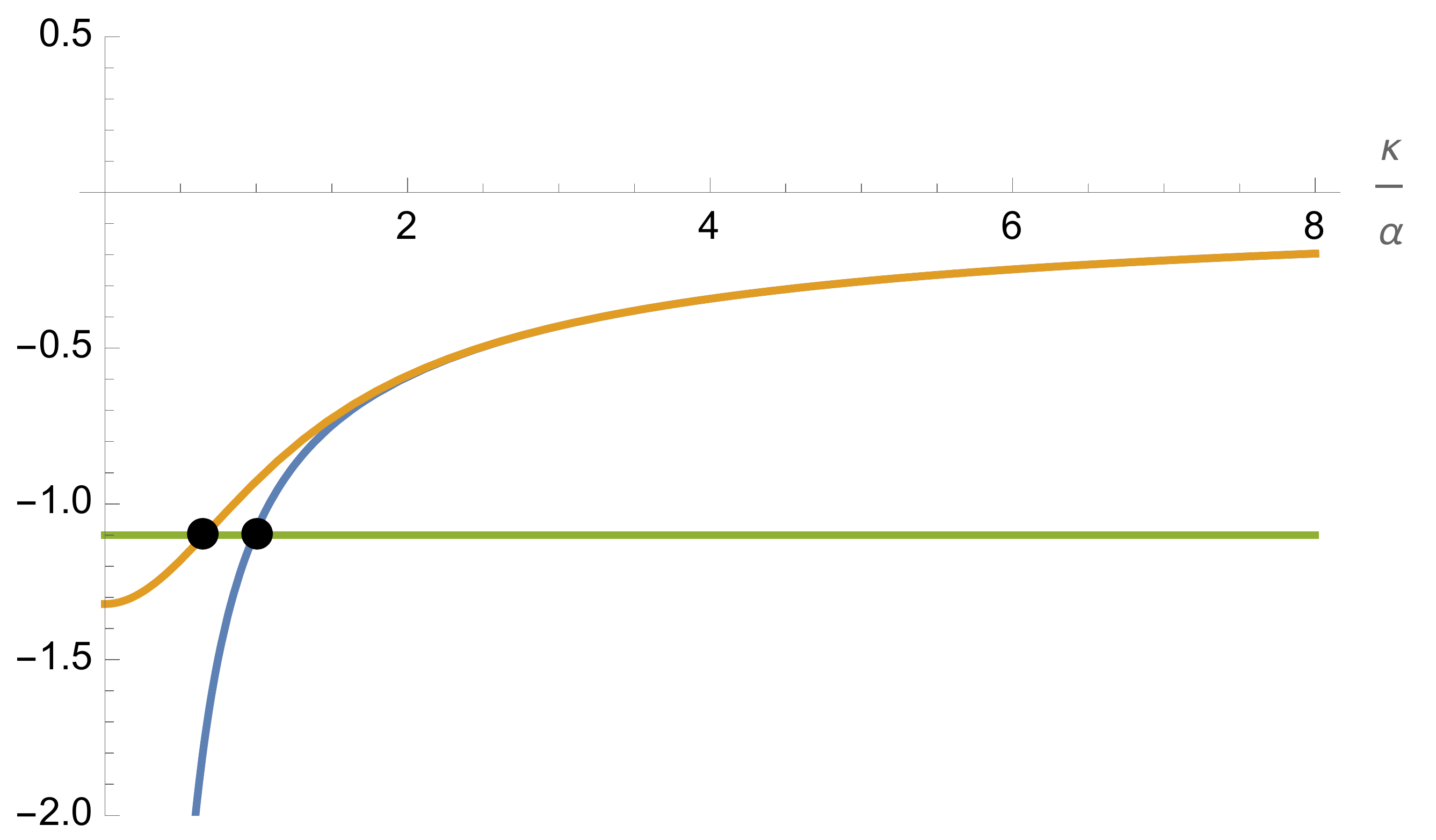}
\label{fig:poco_2}
}
\caption{Graphical solution of the energy spectrum of the double well for $\nu_s=0.6$, $\nu_c=0.4$, and $\varepsilon(\alpha L_{\nu_c})^{2\nu_c}=-1.1$. The spectrum is the black points where the blue curves (even modes given by the RHS of equation (\ref{k_UV_L})) and beige curves (odd modes given by the RHS of equation (\ref{k_IR_L})) touch the green line, $\varepsilon(\alpha L_{\nu_c})^{2\nu_c}=-1.1$. (a) Positive energies, $k>0$. (b) Negative energies, $k\rightarrow i\kappa$, with $\kappa>0$.
}
\label{fig:espectro_IR_L}
\end{figure}

Now we are in the pink square of Figure \ref{fig:esp_g}, with $\alpha^{2\nu_c}\frac{B_{\nu_c}(k)}{A_{\nu_c}(k)}=\varepsilon(\alpha L)^{2\nu_c}$. Equation (\ref{master_2}) provides for the even eigenstates ($A_{\nu_s}(k)=0$)
\begin{eqnarray}
\varepsilon(\alpha L_{\nu_c})^{2\nu_c}=\frac{\Gamma(\nu_c)}{\Gamma(-\nu_c)}\frac{\Gamma\left(\frac{1}{2}-\frac{\nu_s}{2}-\frac{\nu_c}{2}+\frac{k}{2\alpha}\right)\Gamma\left(\frac{1}{2}-\frac{\nu_s}{2}-\frac{\nu_c}{2}-\frac{k}{2\alpha}\right)}{\Gamma\left(\frac{1}{2}-\frac{\nu_s}{2}+\frac{\nu_c}{2}+\frac{k}{2\alpha}\right)\Gamma\left(\frac{1}{2}-\frac{\nu_s}{2}+\frac{\nu_c}{2}-\frac{k}{2\alpha}\right)}.\label{k_UV_L}
\end{eqnarray}
The odd eigenstates ($B_{\nu_s}(k)=0$) are given by
\begin{eqnarray}
\varepsilon(\alpha L_{\nu_c})^{2\nu_c}=\frac{\Gamma(\nu_c)}{\Gamma(-\nu_c)}\frac{\Gamma\left(\frac{1}{2}+\frac{\nu_s}{2}-\frac{\nu_c}{2}+\frac{k}{2\alpha}\right)\Gamma\left(\frac{1}{2}+\frac{\nu_s}{2}-\frac{\nu_c}{2}-\frac{k}{2\alpha}\right)}{\Gamma\left(\frac{1}{2}+\frac{\nu_s}{2}+\frac{\nu_c}{2}+\frac{k}{2\alpha}\right)\Gamma\left(\frac{1}{2}+\frac{\nu_s}{2}+\frac{\nu_c}{2}-\frac{k}{2\alpha}\right)}.\label{k_IR_L}
\end{eqnarray}
The right-hand sides of the above equations remain real and valid even for negative energies ($k\rightarrow i\kappa$). In Figure \ref{fig:espectro_IR_L}, we have the curves that provide the first energy states for $\nu_s=0.6$, $\nu_c=0.4$ and $\varepsilon(\alpha L_{\nu_c})^{2\nu_c}=-1.1$ ($\varepsilon=-1$). The spectrum, for both positive and negative energies, is given by the black points where the blue curves (even modes) and the beige curves (odd modes) touch the green line, the value of $\varepsilon(\alpha L_{\nu_c})^{2\nu_c}$. The following conclusions (valid even for other parameter values) are evident from analyzing Figures \ref{fig:poco_1} and \ref{fig:poco_2}: i) the spectrum alternates indefinitely between even and odd modes; ii) the ground state is always an even eigenstate.

\begin{figure}[ht] 
\centering
\subfigure[]{
\includegraphics[scale=0.5]{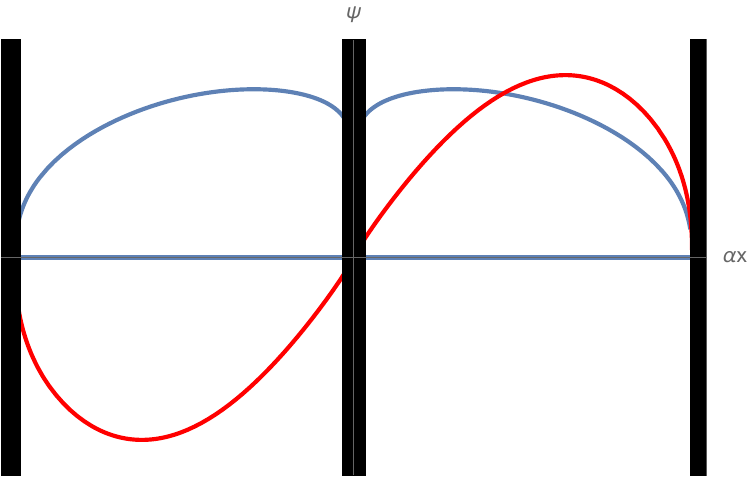}
\label{fig:psi_2_a}
}
\subfigure[]{
\includegraphics[scale=0.5]{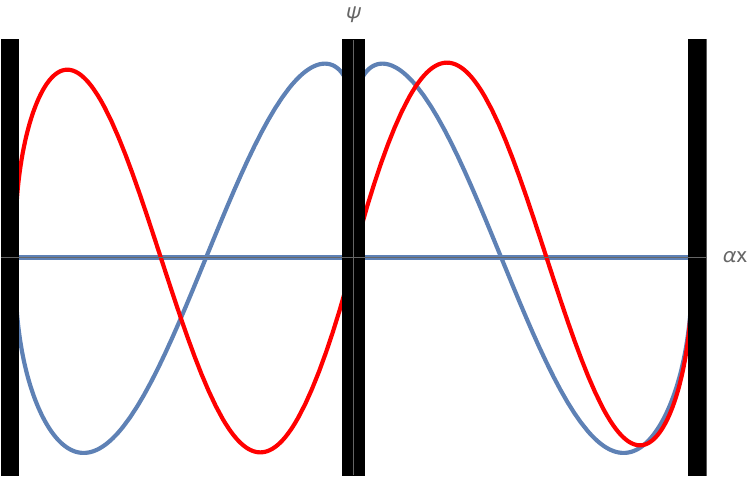}
\label{fig:psi_2_b}
}
\caption{Unnormalized energy eigenstates of the double well associated with $\nu_s=0.4$ and $\nu_c=0.1$ ($-1/4<g_c<3/4$): (a) Ground state in blue and first excited state in red. (b) The second excited state is in blue, and the third excited state is in red. The black bands represent the regions $|x|<R$ and $|\frac{\pi}{2\alpha}-x|<R$.
}
\label{fig:duplo_IR}
\end{figure}

When $L_{\nu_c}=0$, equations (\ref{k_UV_L}) and (\ref{k_IR_L}), as expected, reproduce (\ref{k_UV_IR}) and (\ref{k_IR_IR}), respectively. The other fixed point, $L_{\nu_c}\rightarrow\infty$, is also an option where the spectrum is
\begin{eqnarray}
&&\frac{k_0^{(even)}}{\alpha}=|1-\nu_s-\nu_c|, \ \frac{k_n^{(even)}}{\alpha}=2n+1-\nu_s-\nu_c, \ n\in\mathbb{Z}_{\ge1},\\
&&\frac{k_n^{(odd)}}{\alpha}=2n+1+\nu_s-\nu_c,\ n\in\mathbb{Z}_{\ge0},
\end{eqnarray}
and, again, the spectrum alternates between even and odd states following $k_{0}^{(even)}<k_0^{(odd)}<k_{1}^{(even)}<k_1^{(odd)}<\ldots$. In Figure \ref{fig:duplo_IR}, we have the eigenfunctions of the first energy modes at the IR$_{\nu_c}$ fixed point for $\nu_s=0.4$ e $\nu_c=0.1$. In Figure \ref{fig:psi_2_a}, the blue curve is the ground state, and the red curve is the first excited state, while in Figure \ref{fig:psi_2_b}, the blue curve is the second excited state, and the red curve is the third. Again, we have well-behaved eigenfunctions that follow the \emph{Node Theorem}.
%%%%%%%%%%%%%%%%%%%%%%%%%%%%%%%%%%%%%%%%%%%%%%%%
%%%%%%%%%%%%%%%%%%%%%%%%%%%%%%%%%%%%%%%%%%%%%%%%t

%%%%%%%%%%%%%%%%%%%%%%%%%%%%%%%%%%%%%%%%%%%%%%%%
%%%%%%%%%%%%%%%%%%%%%%%%%%%%%%%%%%%%%%%%%%%%%%%%t 
\subsubsection{Exceptional case $g_c=g_s$ ($\nu_c=\nu_s=\nu$)}\label{sec:part}

\begin{figure}[ht] 
\centering
\subfigure[]{
\includegraphics[scale=0.5]{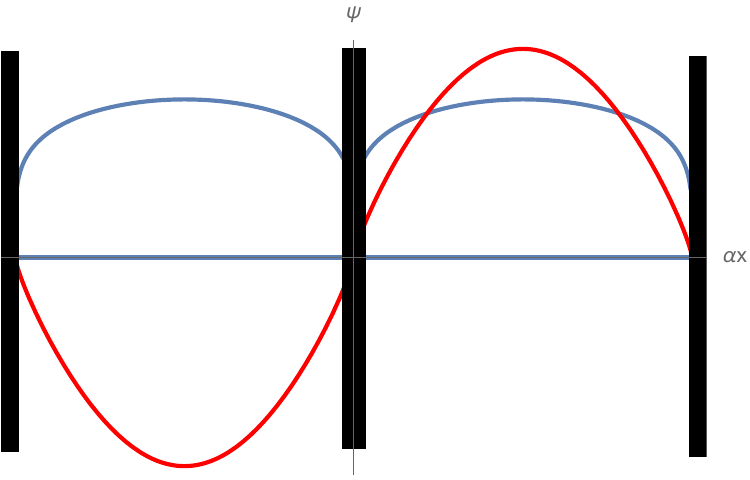}
\label{fig:g_g_1_a}
}
\subfigure[]{
\includegraphics[scale=0.5]{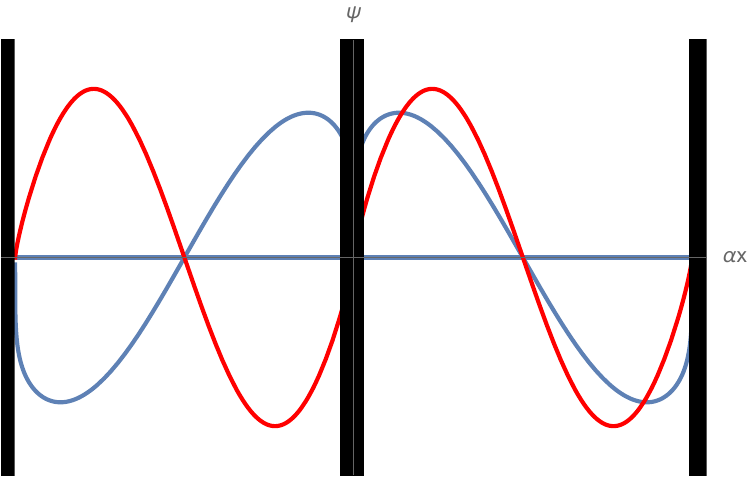}
\label{fig:g_g_1_b}
}
\subfigure[]{
\includegraphics[scale=0.5]{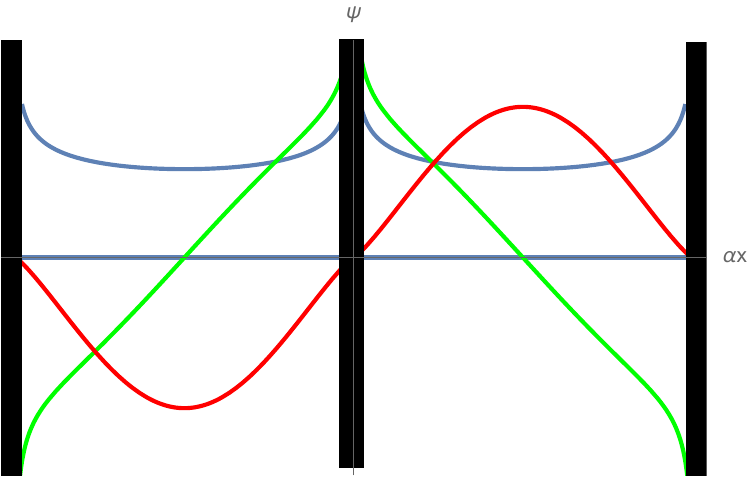}
\label{fig:g_g_2_a}
}
\subfigure[]{
\includegraphics[scale=0.5]{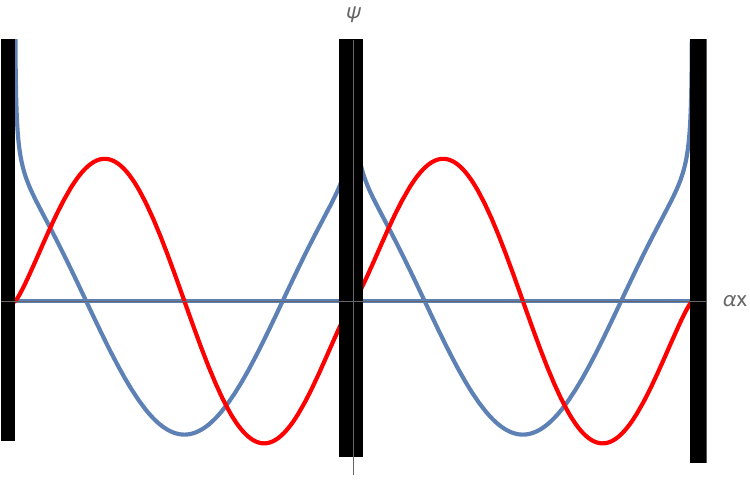}
\label{fig:g_g_2_b}
}
\caption{Unnormalized energy eigenstates of the double well: (a) Ground state in blue and first excited state in red for $\nu_s=\nu_c=0.3$. (b) The second excited state is in blue, and the third excited state is in red for $\nu_s=\nu_c=0.3$. (c) Ground state in blue, first excited state in green, and second excited state in red for $\nu_s=\nu_c=0.7$. (d) The third excited state is in blue, and the fourth excited state is in red for $\nu_s=\nu_c=0.7$. The black bands represent the regions $|x|<R$ and $|\frac{\pi}{2\alpha}-x|<R$.
}
\label{fig:g_g}
\end{figure}

The coincidence in the values of the two couplings, the white line on the diagonal of Figure \ref{fig:esp_g}, can be analyzed directly from equations (\ref{k_UV_L}) and (\ref{k_IR_L}), without the need for a separate study. The exception, our subject of interest, occurs when the fixed points $L_{\nu_c}=0$ and $L_{\nu_c}\rightarrow\infty$ are selected. In this case, some of the poles and zeros of (\ref{k_UV_L}) and (\ref{k_IR_L}) become spurious \cite{Landau_3}, in particular, equation (\ref{k_UV_L}) no longer has zeros corresponding to bound states, only poles, while the opposite occurs for (\ref{k_IR_L}). Thus, it is impossible to construct even modes for $L_{\nu_c}=0$ and odd modes for $L_{\nu_c}\rightarrow\infty$. An interesting solution is to take $L_{\nu_c}\rightarrow\infty$ in (\ref{k_UV_L}) to form the even modes ($A_{\nu_s}(k)=0$, resulting in the spectrum
\begin{eqnarray}
&&\frac{k_0^{(even)}}{\alpha}=|1-2\nu|, \ \frac{k_n^{(even)}}{\alpha}=2n+1-2\nu, \ n\in\mathbb{Z}_{\ge1},\label{k_g_g_p}
\end{eqnarray}
and $L_{\nu_c}=0$ in (\ref{k_IR_L}) to generate the odd modes ($B_{\nu_s}(k)=0$), whose spectrum is
\begin{eqnarray}
\frac{k_n^{(odd)}}{\alpha}=2n+1+2\nu, \ n\in\mathbb{Z}_{\ge0}.\label{k_g_g_i}
\end{eqnarray}
When $0<\nu<1/2$, the spectrum ordering is the usual $k_{0}^{(even)}<k_{0}^{(odd)}<k_{1}^{(even)}<k_{1}^{(odd)}<\ldots$, obeying the \emph{Node Theorem} - as shown in Figures \ref{fig:g_g_1_a} and \ref{fig:g_g_1_b} for $\nu=0.3$. However, when $1/2<\nu<1$, an ``anomaly'' occurs in the ordering: $k_{0}^{(even)}<k_{1}^{(even)}<k_{0}^{(odd)}<k_{1}^{(even)}<k_{1}^{(odd)}<\ldots$, meaning the first two modes are even, and only after that do we have the usual alternation between even and odd modes. The \emph{Node Theorem} is also not valid, as the ``anomaly'' creates a phase shift in the number of nodes of each eigenstate, e.g., $\psi_{0,\nu}(x)$ has zero nodes, $\psi_{1,\nu}(x)$ has two nodes, $\psi_{2,\nu}(x)$ has one node, $\psi_{3,\nu}(x)$ has four nodes, $\psi_{4,\nu}(x)$ has three nodes. That is, excluding the ground state, we have the following rule: $\psi_{2n}(x)$ has $2n-1$ nodes, and $\psi_{2n-1}$ has $2n$ nodes, $n\in\mathbb{Z}_{\ge1}$. In Figures \ref{fig:g_g_2_a} and \ref{fig:g_g_2_b}, we have the first eigenfunctions for $\nu=0.7$.
%%%%%%%%%%%%%%%%%%%%%%%%%%%%%%%%%%%%%%%%%%%%%%%%
%%%%%%%%%%%%%%%%%%%%%%%%%%%%%%%%%%%%%%%%%%%%%%%%t 

\subsection{$g_c<-1/4$ ($\nu_c\rightarrow i|\nu_c|$)}
\begin{figure}[ht] 
\centering
\subfigure[]{
\includegraphics[scale=0.15]{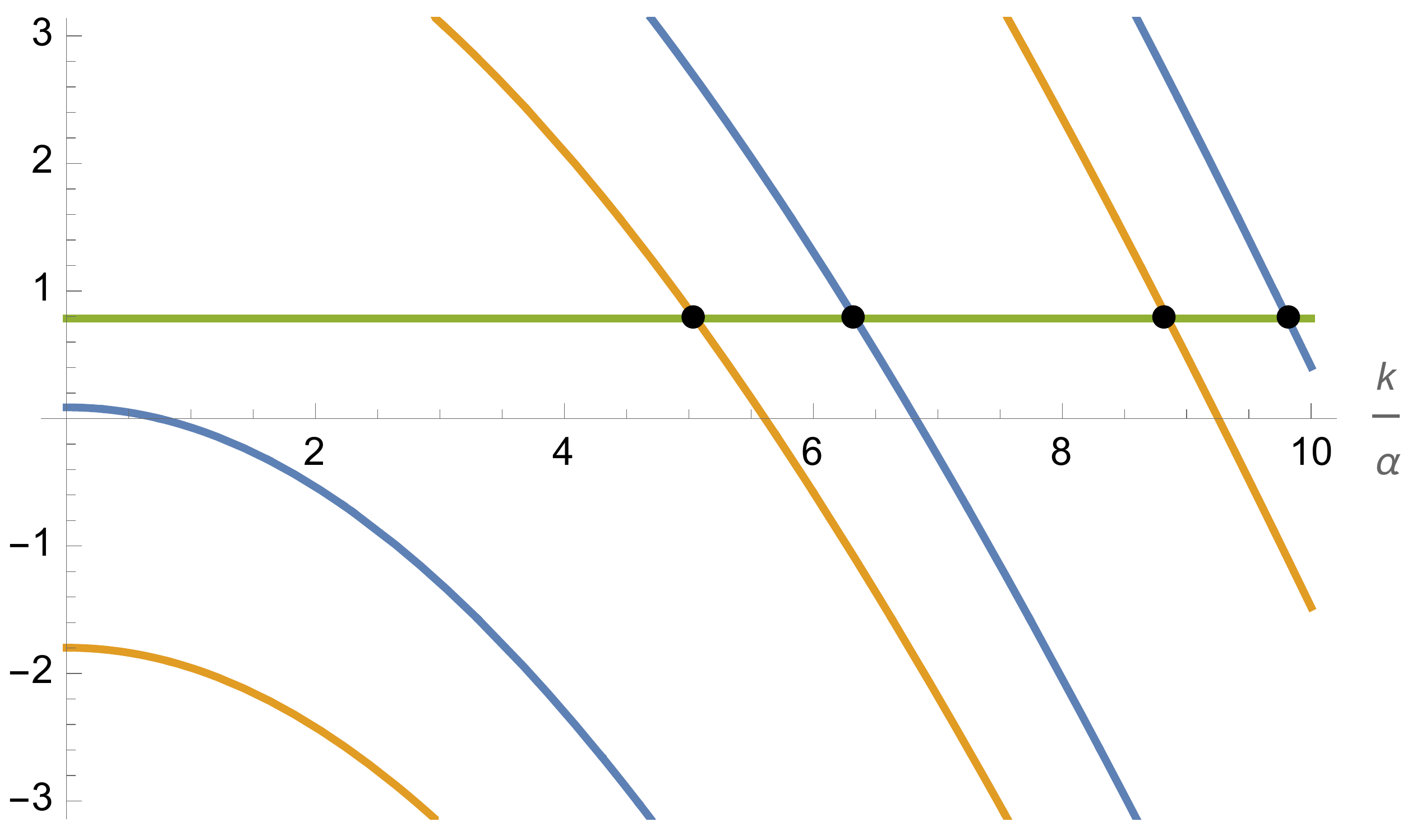}
\label{fig:poco_forte_p}
}
\subfigure[]{
\includegraphics[scale=0.15]{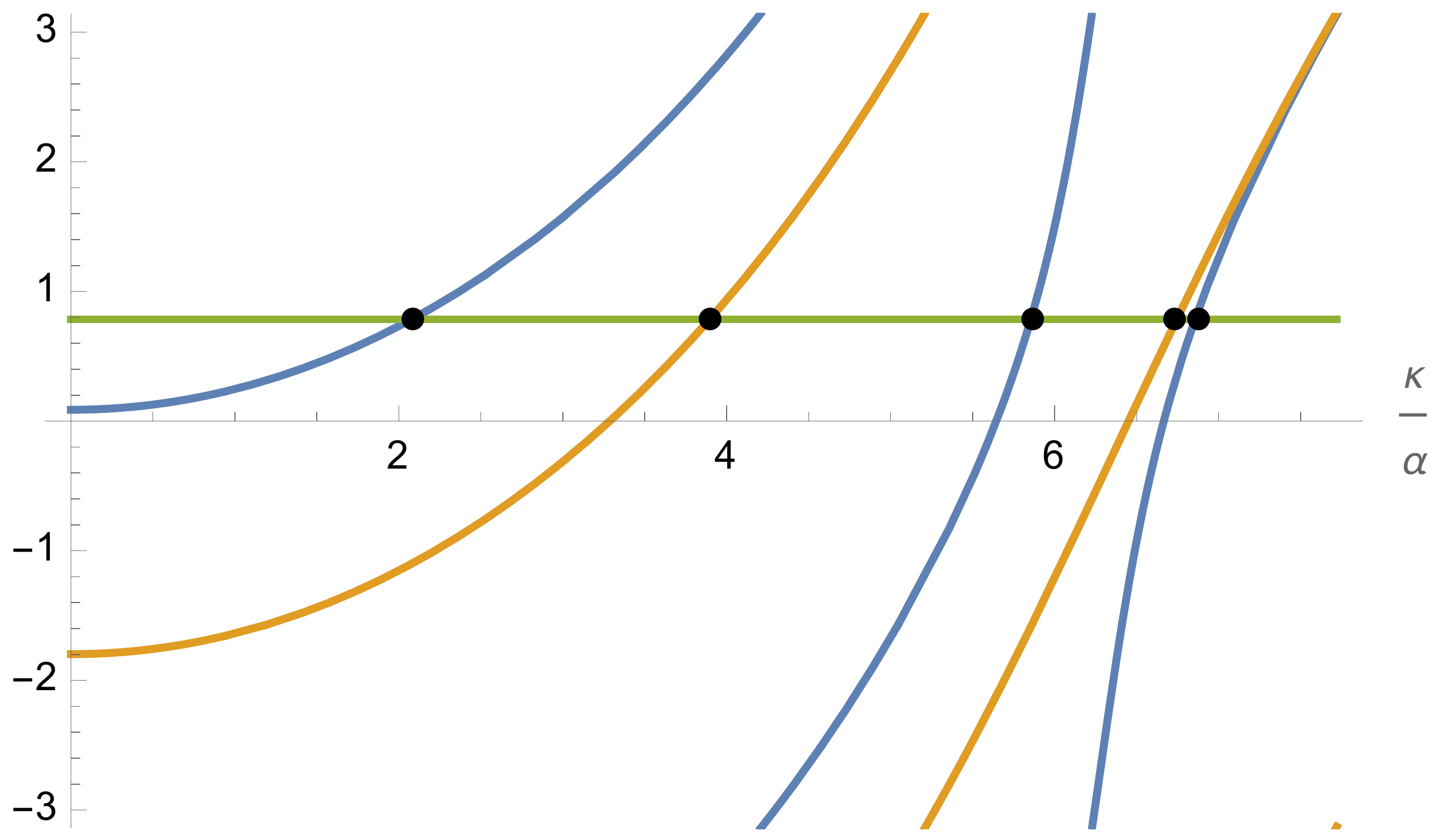}
\label{fig:poco_forte_n}
}
\caption{Graphical solution of the energy spectrum of the double well for $\nu_s=0.7$, $|\nu_c|=2\pi$ and $\theta_{\nu_c}=\frac{\pi}{4}$. In the Figure, the spectrum is the black points where the blue curves (even modes provided by the RHS of equation (\ref{k_UV_im})) and beige curves (odd modes given by the RHS of equation (\ref{k_IR_im})) touch the green line, representing $\varepsilon(\alpha L_{\nu_c})^{2\nu_c}=-1.1$. (a) Positive energies, $k>0$. (b) Negative energies, $k\rightarrow i\kappa$, with $\kappa>0$.
}
\label{fig:poco_forte}
\end{figure}

The last case corresponds to the strong-attractive regime at the ends of the well, represented by the yellow part on the left side of Figure \ref{fig:esp_g}. The spectrum of the even modes is determined by
\begin{eqnarray}
e^{-i\theta_{\nu_c}}=-\frac{\Gamma(i|\nu_c|)}{\Gamma(-i|\nu_c|)}\frac{\Gamma\left(\frac{1}{2}-\frac{\nu_s}{2}-i\frac{|\nu_c|}{2}+\frac{k}{2\alpha}\right)\Gamma\left(\frac{1}{2}-\frac{\nu_s}{2}-i\frac{|\nu_c|}{2}-\frac{k}{2\alpha}\right)}{\Gamma\left(\frac{1}{2}-\frac{\nu_s}{2}+i\frac{|\nu_c|}{2}+\frac{k}{2\alpha}\right)\Gamma\left(\frac{1}{2}-\frac{\nu_s}{2}+i\frac{|\nu_c|}{2}-\frac{k}{2\alpha}\right)}.\label{k_UV_im}
\end{eqnarray}
and the odd states have energies that satisfy
\begin{eqnarray}
e^{-i\theta_{\nu_c}}=-\frac{\Gamma(i|\nu_c|)}{\Gamma(-i|\nu_c|)}\frac{\Gamma\left(\frac{1}{2}+\frac{\nu_s}{2}-i\frac{|\nu_c|}{2}+\frac{k}{2\alpha}\right)\Gamma\left(\frac{1}{2}+\frac{\nu_s}{2}-i\frac{|\nu_c|}{2}-\frac{k}{2\alpha}\right)}{\Gamma\left(\frac{1}{2}+\frac{\nu_s}{2}+i\frac{|\nu_c|}{2}+\frac{k}{2\alpha}\right)\Gamma\left(\frac{1}{2}+\frac{\nu_s}{2}+i\frac{|\nu_c|}{2}-\frac{k}{2\alpha}\right)}.\label{k_IR_im}
\end{eqnarray}
Both equations are valid for positive energies ($k>0$) as well as negative energies ($k=i\kappa$, $\kappa>0$). In Figure \ref{fig:poco_forte}, we have the graphical determination of the first modes for $\theta_c=\pi/4$, $\nu_s=0.7$ and $|\nu_c|=2\pi$. For deep negative modes, $\kappa/\alpha\gg1$, the parameter $\nu_s\in(0,1)$ becomes irrelevant, and the even and odd modes tend to degenerate, with the ratio $\kappa_{n+1}/\kappa_n$ tending to the value $e^{\frac{\pi}{|\nu_c|}}$, consistent with the \emph{Efimov Rule}.
%%%%%%%%%%%%%%%%%%%%%%%%%%%%%%%%%%%%%%%%%%%%%%%%
%%%%%%%%%%%%%%%%%%%%%%%%%%%%%%%%%%%%%%%%%%%%%%%%t

\section{Discussion}\label{sec:conc}
In this article, we systematically study the renormalization procedure for the one-dimensional stationary Schr\"odinger equation with a potential exhibiting two inverse-square singularities. The method was applied to the trigonometric P\"oschl-Teller potential, defined between two singularities of this type. The renormalized energy spectrum was determined across the entire $g_s-g_c$ coupling plane, generating various regions with distinct properties, as shown in Figure \ref{fig:esp_g}. The fact that the TPT potential is SUSY and shape-invariant simplifies the analysis, and we have to fix Dirichlet boundary conditions at both singularities when $g_s,g_c\ge-1/4$, as illustrated in Figure \ref{fig:esp_g_2}, showing that there is no need for a self-adjoint extension in this region of the coupling space, as done in \cite{Ishibashi:2004wx}. On the other hand, when at least one of the couplings is in the strong-attractive region, $g<-1/4$, the naive use of the boundary conditions leads to an imaginary spectrum, so the Hamiltonian is not self-adjoint. In these cases, renormalization is necessary and generates a family of eigenstates and eigenvalues dependent on one or two parameters, which introduce anomalous scales via dimensional transmutation, breaking the asymptotic conformal symmetry, \cite{Camblong:2000qn,Kaplan:2009kr}. The spectrum is discrete but without a lower bound since at least one coupling is on the \emph{strong-attractive} regime. In general, the bound states are determined by complicated transcendental relations with graphical solutions, but for deep states, where $\kappa/\alpha\gg1$, they always converge to the \emph{Efimov Rule}\footnote{The exception occurs when $g_s,g_c<-1/4$, where there is competition between the two singularities, and we were unable to identify a pattern in the spectrum.}, $\kappa_{n+1}/\kappa_n=e^{\frac{\pi}{|\nu|}}$. That means that deep bound states have wave functions localized near $x=x_s$, so the potential is effectively conformal, $\frac{2m}{\hbar^2}V(x)= g/(x-x_s)^2$.

In our previous work, \cite{CAMARADASILVA2024169549z}, a similar analysis was carried out, where we performed the renormalization of the hyperbolic P\"oschl-Teller (HPT) potential. Although the HPT potential has only one singularity, it gives rise to a more wealthy spectral behavior than the TPT one. Apart from the usual \emph{standard bound states}, its renormalization generates specific anti-bound states and resonance modes, \cite{CAMARADASILVA2024169549z,QNM_BH_Julio}. Its applications concern mainly the description of some black hole quasi-normal modes spectra. \cite{PhysRevD63124015,cardona_2017_PT,Molina:2003ff,Churilova:2021nnc}. The HPT and TPT models are formally connected by the mapping $\alpha \rightarrow i\alpha$. This correspondence does not reflect the renormalization process, as it introduces a second singularity that modifies the spatial range of the TPT model. Thus, the HPT continuous spectrum, anti-bound states, and resonance modes, characterized by at most one parameter, are destroyed, and the TPT model has only bound states. On the other hand, the presence of two singularities implies that the two independent renormalization processes lead to a richer structure with an energy spectrum that can depend on one or two parameters.

The renormalization of the TPT model presented in this article has numerous applications to modern theoretical physics problems. A specific case we are interested in is the dynamics of free scalar, vector, and tensor fields in $AdS_d$ space. They are all described by TPT potential in global coordinates, \cite{Ishibashi:2004wx}. In this context, the renormalized modes lie below the BF bound ($g<-1/4$) and are excluded (they are unstable). On the other hand, the extension of the potential to a double well, as done in Section \ref{sec:duplo}, opens a new possibility: $AdS_d$ ($d>2$) space is a hemisphere with the boundary being the equator, and the double well can be view as an extended $AdS_d$, essentially two $AdS_d$ spaces forming a sphere sharing the boundary at the equator. In general, this duplication of the space does not imply anything new, as the spaces are disconnected. However, if the effective mass of the field, $m$, lies in the range $-(d-1)^2/4<\frac{m^2}{\alpha^2}<(3-d(d-2))/4$ ($1/\alpha$ is the $AdS_d$ radius), we have the situation described in Section \ref{sec:duplo}, where $x=0$ is the boundary of both $AdS_d$ spaces (the equator of the sphere) and $x=\pm\pi/2\alpha$ are the poles of the sphere. Then, this field exists in both geometries, and its spectrum diverges from that of a field defined in a single $AdS_d$. In addition to analyzing this promising point in more detail, we intend to study the changes generated by this extended-spectrum in the holographic $CFT_{d-1}$.

Another application of interest concerns the modification of TPT potential by extending its spatial interval when both couplings are of the weak-medium type and are equal. Now, the argument for extending the domain of the wave function can be applied to both singularities and repeated indefinitely. Hence, we have the model on the entire real line as a crystal lattice with a periodic and singular potential\footnote{Our preliminary results suggest that this \emph{lattice potential} has an exact and non-trivial band spectrum.}. Within Anti-de Sitter space-time, such a construction is suitable for dimension $d=2$, where the space-time is a strip with the boundaries being two lines ($x=0$ and $x=\pi/2\alpha$). In the extended version, the new space would be the entire plane, accessible to fields with effective mass in the range $-1/4<\frac{m^2}{\alpha^2}<3/4$ ($1/\alpha$ is the $AdS_2$ radius).

The results of the renormalization of HPT and TPT models, as well as the extended TPT model having indefinitely many inverse-square singularities, contribute to a broader program designed for the renormalization of supersymmetric and shape-invariant potentials without, \cite{dutt1988} and with $\hbar$-dependence, \cite{quesne2008,bougie2010,Odake2009,Odake2010,Odake2011}.

\vspace{0,5cm}

\textbf{Acknowledgments}

I would like to thank GM Sotkov for the manuscript's critical reading and stimulating conversation.

\appendix
\section{Appendix A: $g_s=-1/4$, $\nu_s=0$}\label{sec:app}

If $\nu_c\ge0$, SUSY implies that the second solution is zero, and we have the same spectrum as in (\ref{E_lig_trig}) with $\nu_s=0$. The novelty arises when $g_c<-1/4$\footnote{Dotted line between the green and yellow regions in Figure \ref{fig:esp_g}.}, as now we cannot discard either of the solutions, one of which exhibits a logarithmic singularity at $x=0$. The general form of the stationary state is
\begin{eqnarray}
&&\psi_{k;0,i|\nu_c|}(x)=\sqrt{\frac{\tan(\alpha x)}{\alpha}}\times\nonumber\\
&&\Bigg\{A_{0}(k)\left(\cos(\alpha x)\right)^{-\frac{k}{\alpha}}\,{}_2F_1\left(\frac{1}{2}-i\frac{|\nu_c|}{2}+\frac{k}{2\alpha},\frac{1}{2}+i\frac{|\nu_c|}{2}+\frac{k}{2\alpha};1;-\tan^2(\alpha x)\right)+\nonumber\\
&&\frac{B_{0}(k)}{2}\mathcal{R}e\Bigg[\frac{\Gamma\left(\frac{1}{2}-i\frac{|\nu_c|}{2}+\frac{k}{2\alpha}\right)\Gamma\left(\frac{1}{2}+i\frac{|\nu_c|}{2}+\frac{k}{2\alpha}\right)}{\Gamma\left(1+\frac{k}{\alpha}\right)}\left(\cos(\alpha x)\right)^{-\frac{k}{\alpha}}\times\nonumber\\
&&\,{}_2F_1\left(\frac{1}{2}-i\frac{|\nu_c|}{2}+\frac{k}{2\alpha},\frac{1}{2}+i\frac{|\nu_c|}{2}+\frac{k}{2\alpha};1+\frac{k}{\alpha};\frac{1}{\cos^2(\alpha x)}\right)\Bigg]\Bigg\}.\label{psi_app}
\end{eqnarray}
The real part of the solution was taken in the second term above, as it is complex because the argument of the hypergeometric function is between $1<1/\cos^2(\alpha x)<\infty$. This procedure is necessary because an energy eigenstate of a one-dimensional bound state is always a real function — its current density is zero.

Using the asymptotic forms of the hypergeometric function, \cite{NIST:DLMF}, we will determine the behavior of (\ref{psi_app}) at the system's boundaries. Starting with the origin, we have near $x=0$
\begin{eqnarray}
\psi_{k;0,i|\nu_c|}(x)\approx A_0(k)x^{1/2}-B_0(k)x^{1/2}\left[\zeta(k;|\nu_c|)+\ln(\alpha x)\right],\nonumber\\
\propto x^{1/2}\left(1-\frac{B_0(k)}{A_0(k)-B_0(k)\xi(k;|\nu_c|)}\ln\left(\alpha x\right)\right),\label{psi_0_app}
\end{eqnarray}
where
\begin{eqnarray}
\xi(k;|\nu_c|)\equiv \gamma_E+\frac{1}{2}\Re e\left(\psi^{(0)}\left(\frac{1}{2}+\frac{k}{2\alpha}+i\frac{|\nu_c|}{2}\right)+\psi^{(0)}\left(\frac{1}{2}+\frac{k}{2\alpha}-i\frac{|\nu_c|}{2}\right)\right)
\end{eqnarray}
with $\gamma_E$ being the Euler constant and $\psi^{(0)}(z)$ the Digamma function, \cite{NIST:DLMF}. For $x\rightarrow\frac{\pi}{2\alpha}$ ($k>0$)
\begin{eqnarray}
\psi_{k;0,i|\nu_c|}(x)\approx A_{i|\nu_c|}(k)\left(\frac{\pi}{2\alpha}-x\right)^{\frac{1}{2}+i|\nu_c|}+B_{i|\nu_c|}(k)\left(\frac{\pi}{2\alpha}-x\right)^{\frac{1}{2}-i|\nu_c|},
\end{eqnarray}
with the definitions
\begin{eqnarray}
&&A_{i|\nu_c|}(k)=\frac{\Gamma(-i|\nu_c|)}{\alpha^{-i|\nu_c|}}\Bigg[A_0(k)\frac{1}{\Gamma\left(\frac{1}{2}-i\frac{|\nu_c|}{2}+\frac{k}{2\alpha}\right)\Gamma\left(\frac{1}{2}-i\frac{|\nu_c|}{2}-\frac{k}{2\alpha}\right)}+\nonumber\\
&&\frac{B_0(k)}{4i}\left(e^{\frac{\pi|\nu_c|}{2}}e^{\frac{-i\pi k}{2\alpha}}\mathcal{X}(k;i|\nu_c|)-e^{\frac{-\pi|\nu_c|}{2}}\left(e^{\frac{-i\pi k}{2\alpha}}\mathcal{X}(k;-i|\nu_c|)\right)^*\right)\Bigg],\\
&&B_{i|\nu_c|}(k)=\frac{\Gamma(i|\nu_c|)}{\alpha^{i|\nu_c|}}\Bigg[A_0(k)\frac{1}{\Gamma\left(\frac{1}{2}+i\frac{|\nu_c|}{2}+\frac{k}{2\alpha}\right)\Gamma\left(\frac{1}{2}+i\frac{|\nu_c|}{2}-\frac{k}{2\alpha}\right)}-\nonumber\\
&&\frac{B_0(k)}{4i}\left(e^{\frac{\pi|\nu_c|}{2}}\left(e^{\frac{-i\pi k}{2\alpha}}\mathcal{X}(k;i|\nu_c|)\right)^*-e^{\frac{-\pi|\nu_c|}{2}}e^{\frac{-i\pi k}{2\alpha}}\mathcal{X}(k;-i|\nu_c|)\right)\Bigg].\label{A_B_app}
\end{eqnarray}
Here, we introduced the auxiliary functions
\begin{eqnarray}
\mathcal{X}(k;|\nu_c|)\equiv\frac{\Gamma\left(\frac{1}{2}+i\frac{|\nu_c|}{2}+\frac{k}{2\alpha}\right)}{\Gamma\left(\frac{1}{2}-i\frac{|\nu_c|}{2}+\frac{k}{2\alpha}\right)}.
\end{eqnarray}

\subsection{Renormalization}
\begin{figure}[ht] 
\centering
\subfigure[]{
\includegraphics[scale=0.55]{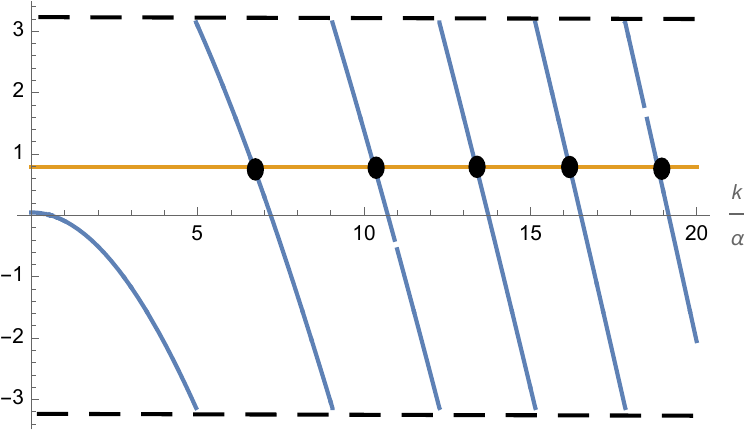}
\label{fig:app1}
}
\subfigure[]{
\includegraphics[scale=0.55]{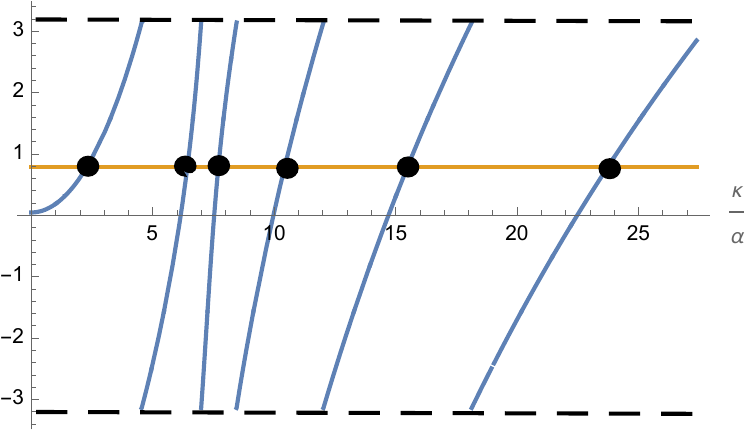}
\label{fig:app2}
}
\caption{Graphical solution of the equation (\ref{app_final}) for $(\mathcal{D},\theta)=(5,\pi/4)$ and $|\nu_c|=7$. The spectrum is the black points where the blue curves ($-i\ln($RHS of \ref{app_final}$)$) touch the beige line ($\theta_{\nu_c}$). (a) Positive energies, $k>0$. (b) Negative energies, $k\rightarrow i\kappa$, with $\kappa>0$.
}
\label{fig:app}
\end{figure}

The asymptotic solution (\ref{psi_0_app}) with $k=0$ must have the form
\begin{eqnarray}
\psi_{0;0,i|\nu_c|}\propto x^{1/2}\left(1-\mathcal{D}\ln(\alpha x)\right),
\end{eqnarray}
The dimensionless parameter $D$ is independent of energy. From this asymptotic, we can calculate
\begin{eqnarray}
R\mathcal{F}(R)=\frac{1}{2}-\frac{\mathcal{D}}{1-\mathcal{D}\ln\left(\alpha R\right)}.
\end{eqnarray}
Repeating the same steps with (\ref{psi_0_app}), $k\neq0$, and comparing the results, we have
\begin{eqnarray}
\mathcal{D}=\frac{B_0(k)}{A_0(k)-B_0(k)\xi(k;|\nu_c|)},
\end{eqnarray}
or alternatively
\begin{eqnarray}
\frac{A_0(k)}{B_0(k)}=\xi(k;|\nu_c|)+\frac{1}{\mathcal{D}},
\end{eqnarray}
The renormalization at $x=\frac{\pi}{2\alpha}$ implies (\ref{ren_g_forte}), so a new parameter, $\theta_{\nu_c}$, is introduced. Since both must be valid, we have the constraint
\begin{eqnarray}
e^{i\theta_{\nu_c}}=-\frac{\Gamma(-i|\nu_c|)}{\Gamma(i|\nu_c|)}\frac{\frac{\left(\xi(k;|\nu_c|)+\frac{1}{\mathcal{D}}\right)}{\Gamma\left(\frac{1}{2}-i\frac{|\nu_c|}{2}+\frac{k}{2\alpha}\right)\Gamma\left(\frac{1}{2}-i\frac{|\nu_c|}{2}-\frac{k}{2\alpha}\right)}-\frac{1}{4i}\left(e^{\frac{\pi|\nu_c|}{2}}e^{\frac{-i\pi k}{2\alpha}}\mathcal{X}(k;i|\nu_c|)-e^{\frac{-\pi|\nu_c|}{2}}\left(e^{\frac{-i\pi k}{2\alpha}}\mathcal{X}(k;-i|\nu_c|)\right)^*\right)}{\frac{\xi(k;|\nu_c|)+\frac{1}{\mathcal{D}}}{\Gamma\left(\frac{1}{2}+i\frac{|\nu_c|}{2}+\frac{k}{2\alpha}\right)\Gamma\left(\frac{1}{2}+i\frac{|\nu_c|}{2}-\frac{k}{2\alpha}\right)}-\frac{1}{4i}\left(e^{\frac{\pi|\nu_c|}{2}}\left(e^{\frac{-i\pi k}{2\alpha}}\mathcal{X}(k;i|\nu_c|)\right)^*-e^{\frac{-\pi|\nu_c|}{2}}e^{\frac{-i\pi k}{2\alpha}}\mathcal{X}(k;-i|\nu_c|)\right)},\label{app_final}
\end{eqnarray}
which discretizes the spectrum (for both signs of energy) in terms of a two-parameter family $(\mathcal{D},\theta_{\nu_c})$. In Figure \ref{fig:app} we present the graphical solution of the equation (\ref{app_final}) for $(\mathcal{D},\theta)=(5,\pi/4)$ and $|\nu_c|=7$. For deep negative energies, $\kappa/\alpha\gg1$, the ratio $\kappa_{n+1}/\kappa_n$ tends to the value $e^{\frac{\pi}{|\nu_c|}}$ - for our example, $\kappa_{11}/\kappa_{10}=1.56604\approx 1.56643=e^{\frac{\pi}{|\nu_c|}}$, $|\nu_c|=7$.
%%%%%%%%%%%%%%%%%%%%%%%%%%%%%%%%%%%%%%%%%%%%%%%%
%%%%%%%%%%%%%%%%%%%%%%%%%%%%%%%%%%%%

We emphasize that $\psi_{k;0,i|\nu_c|}(x)$ is real, and this implies that $B_{i|\nu|}(k)=\left(A_{i|\nu|}(k)\right)^*$ in equation (\ref{A_B_app}), a fact evident for positive energies $k>0$ since $\Sigma(k;-i\nu_c)=\left(\Sigma(k;i\nu_c)\right)^*$. The same conclusion is not trivial for negative energies, $k\rightarrow i\kappa$ ($\kappa>0$). In these cases, we must be careful when using our results when it is necessary to take the complex conjugate; first, we perform the substitution $k\rightarrow i\kappa$, and only then the conjugation, e.g.
\begin{eqnarray*}
\left(e^{\frac{-i\pi k}{2\alpha}}\mathcal{X}(k;i|\nu_c|)\right)^*\Big|_{k\rightarrow i\kappa}=\left(e^{\frac{\pi \kappa}{2\alpha}}\mathcal{X}(i\kappa;i|\nu_c|)\right)^*=e^{\frac{\pi \kappa}{2\alpha}}\mathcal{X}(-i\kappa;-i|\nu_c|).
\end{eqnarray*}

\bibliographystyle{utphys}
\bibliography{PT_trig_generalizado}

\end{document}